\documentclass[11pt,journal,comsoc,twocolumn,twoside]{IEEEtran}

\usepackage{graphicx,cite,epsfig,amssymb,amsmath}
\usepackage{amsthm}
\newcommand{\halfcheck}{%
  \ooalign{\checkmark\cr\hidewidth\raisebox{.5ex}{--}\hidewidth}%
}

\makeatletter

\usepackage{cite}\usepackage{subfigure}\usepackage{stfloats}\@ifundefined{definecolor}
 {\usepackage{color}}{}
\usepackage{pifont}\usepackage{latexsym}\usepackage{algorithm}\usepackage{algorithmic}

\newtheorem{remark}{\textbf{Remark}}

\makeatother

\begin{document}
\title{Beyond 1$\rightarrow$N Decoding: Capacity-Aware Rateless Polar Codes for IR-HARQ}

\author{Huazi~Zhang,
        Xianbin~Wang,
        Jiajie~Tong,
        Jun~Wang,
        Wen~Tong,
    \thanks{All authors are with Huawei Technologies Co. Ltd.
    (Email: \{zhanghuazi, wangxianbin1, tongjiajie, justin.wangjun, tongwen\}@huawei.com).}
}
\maketitle

\markboth{Zhang, Wang and Tong: Beyond 1$\rightarrow$N Decoding: Capacity-Aware Rateless Polar Codes for IR-HARQ}{Zhang, Wang and Tong: Beyond 1$\rightarrow$N Decoding: Capacity-Aware Rateless Polar Codes for IR-HARQ}

\begin{abstract}
This paper introduces a novel framework for polar codes, designed for flexible Incremental Redundancy Hybrid Automatic Repeat Request (IR-HARQ). By generalizing the decoding order beyond the standard $1 \rightarrow N$ sequence, we enable a capacity-aware scheduling strategy that prioritizes the decoding of reliable subblocks. The framework integrates nested parity-check polar construction and reverse bit-mapping to support continuous and arbitrary transmission lengths $E \in [N_{\min}, N_{\max}]$. Simulation results show that the proposed rateless codes match the coding gain of independently optimized fixed-rate codes across the entire range of rates and lengths. With a validated hardware implementation, this work provides a practical solution for next-generation wireless data channels.
\end{abstract}


\IEEEpeerreviewmaketitle

\section{Introduction}\label{section:intro}
\subsection{Motivation}\label{section:intro:motivation}
Hybrid automatic repeat request (HARQ) combines forward error correction with retransmissions to improve spectral efficiency and reliability on wireless data channels. Incremental redundancy HARQ (IR-HARQ) transmits new parity bits in each retransmission instead of simply repeating the same codeword, which significantly increases spectral efficiency and robustness~\cite{RCPC(IR-HARQ)}. IR-HARQ integrates tightly with link adaptation and enables wireless systems to operate efficiently over a wide range of SNRs and mobility conditions.

\subsection{Rateless IR-HARQ}\label{section:intro:ratelessIR}
The rateless IR-HARQ property is essential to modern wireless communication systems. It ensures that transmission rates are always aligned with the instantaneous channel capacity without requiring complex, fixed-rate code constructions.

Consider a code block of $K$ information bits to be encoded once into a mother codeword of length $N_{\max}$ bits (the maximum amount of coded redundancy available for that block). The $N_{\max}$ coded bits are written sequentially into a circular buffer.

At each (re)transmission, the transmitter does not re-encode. Instead, it reads out $E_t$ bits from the circular buffer (where $E_t$ can vary arbitrarily from transmission to transmission, depending on the scheduled resources at time $t$) and sends them over the channel. When the end of the buffer is reached, reading continues from the beginning (wrap around). The total number of coded bits transmitted after $T$ transmissions is
\begin{equation*}
 E^{(T)} = \sum_{t=1}^{T} E_t.
\end{equation*}

The effective code rate after combining all received redundancy up to $T$ transmissions is then
\begin{equation*}
R^{(T)} = \frac{K}{E^{(T)}}.
\end{equation*}

Because $E_t$ and the number of transmissions $T$ are not fixed in advance, the scheme is rateless: the effective code rate is not predetermined at encoding~\cite{LTcodes(Rateless)}.

\begin{figure}
    \centering
    \includegraphics[width=\linewidth]{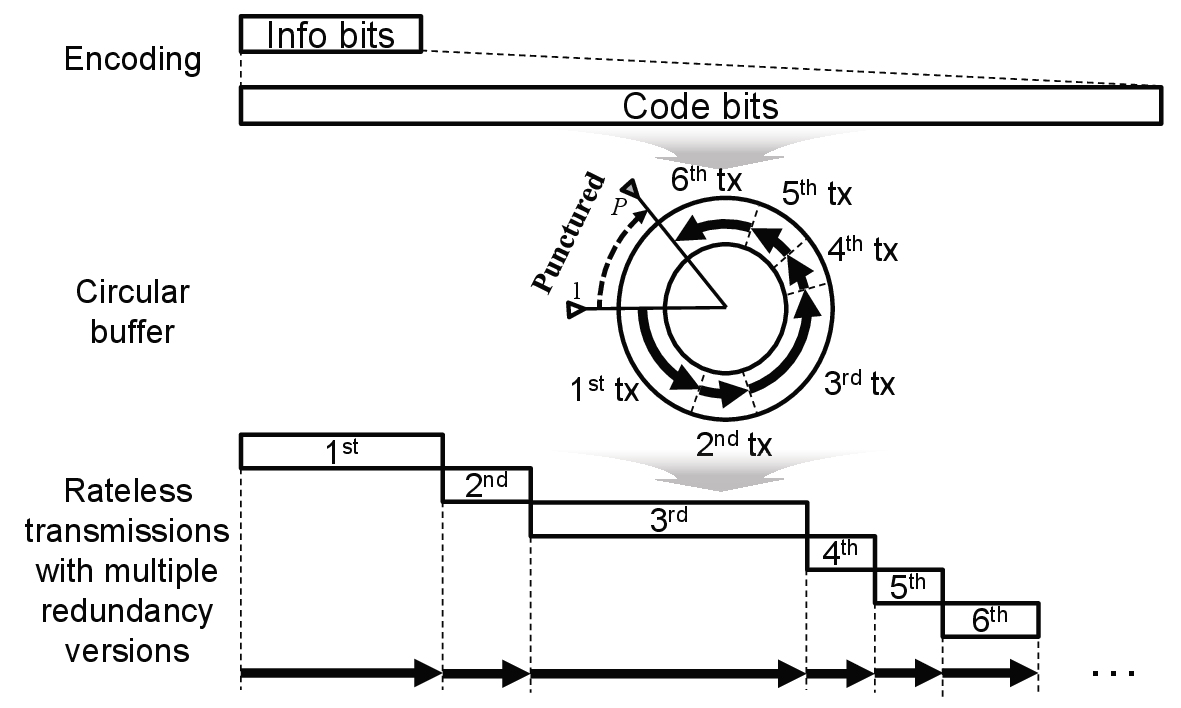}
    \caption{Rateless IR-HARQ}
    \label{fig:Rateless-IR-HARQ}
\end{figure}

The rateless IR-HARQ scheme, as illustrated in Fig.~\ref{fig:Rateless-IR-HARQ} is required to achieve
\begin{enumerate}
  \item \emph{Fine-granularity, on-demand rate adaptation (flexibility)}: support fine-granularity incremental redundancy with arbitrary $E_t$ and arbitrary $T$, allowing an essentially continuous set of effective code rates $R^{(T)} = K/E^{(T)}$ (up to the limit $E^{(T)} \leq N_{\max}$);
  \item \emph{Rate-compatibility / nested codewords}: any lower-rate codeword is an extension of any higher-rate one. The family of codes $\{ \mathcal{C}(K, E) : K \leq E \leq N_{\max} \}$ must be nested (rate-compatible) in the sense that if $E^{(1)} < E^{(2)}$, then the length-$E^{(1)}$ codeword is the prefix (or a fixed-index subset) of the length-$E^{(2)}$ codeword;
  \item \emph{Near-optimal performance at every effective rate (no loss from nesting)}: the performance at rate $R^{(T)}$ is not significantly worse than that of an independently designed, non-rateless code of the same rate and blocklength.
\end{enumerate}

\section{Existing IR-HARQ schemes in 3GPP standards}\label{section:existing3GPP}
In this section, we briefly review existing IR-HARQ schemes deployed from 3G to 5G, and discuss whether they have fully realized the rateless IR-HARQ capability.
\subsection{3G/4G Turbo codes}\label{section:existing3GPP:turbo}
IR-HARQ was first widely deployed in 3G systems~\cite{3G}, where turbo codes~\cite{Turbo} with rate-compatible puncturing support adaptive transmission on data channels by starting from a relatively high-rate turbo code and progressively sending additional parity bits upon NACKs (negative-acknowledgments). This mechanism was reused in LTE (4G)~\cite{4G}, still based on turbo codes but employing a unified circular-buffer rate-matching structure that defined four redundancy versions, i.e., $RV = {0,1,2,3}$.

\begin{figure}
    \centering
    \includegraphics[width=\linewidth]{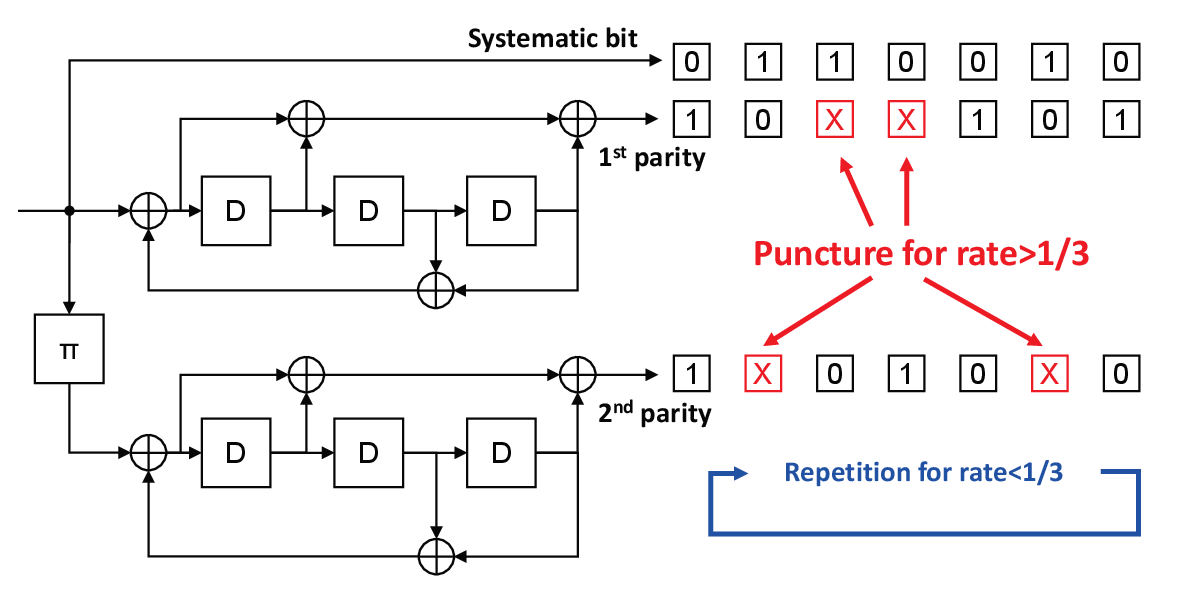}
    \caption{Turbo-based IR-HARQ (LTE)}
    \label{fig:Turbo-IR-HARQ}
\end{figure}

As shown in Fig.~\ref{fig:Turbo-IR-HARQ}, the LTE turbo code is based on two identical recursive systematic convolutional (RSC) component encoders concatenated in parallel through an interleaver. Each information bit is transmitted systematically and is protected by one parity bit from each component encoder. As a result, the ``mother'' code naturally has a nominal code rate of $1/3$. LTE turbo codes, including the code distance and decoding threshold, are optimized to achieve near-optimal performance around this mother code of $1/3$.

To support higher code rates in IR-HARQ, LTE employs extensive puncturing of the parity streams to effectively increase the rate above $1/3$. However, at these high rates the resulting punctured turbo codes deviate substantially from the properties of the underlying mother code. Consequently, turbo codes exhibit significant performance loss for high code rates, e.g., error floors due to low-weight codewords. On the other hand, for code rates below $1/3$, extra redundancy is offered by simple repetition rather than by generating new, independent parity symbols. It only provides additional energy gain.

Since turbo codes do not provide near-optimal performance at every effective rate, turbo-based IR-HARQ cannot fully exploit the potential of rateless IR-HARQ.

\subsection{5G LDPC codes}\label{section:existing3GPP:LDPC}
Starting with 5G NR~\cite{5G}, LDPC codes~\cite{LDPC:Gallager} are adopted for data channels. The NR LDPC design~\cite{5G} includes a raptor-like extension mechanism that can, in principle, generate unlimited parity bits (in practice constrained by the minimum mother code rate). This enables fine-grained incremental redundancy while exploiting the quasi-cyclic structure of LDPC codes for efficient implementation.

In legacy systems such as WiFi, LDPC codes are defined only for a finite set of fixed code rates and block lengths. Each codeword is generated from a parity-check matrix tailored to a specific rate. This inherently limits the granularity of incremental redundancy and makes it difficult to support truly rateless IR-HARQ.

\begin{figure}
    \centering
    \includegraphics[width=\linewidth]{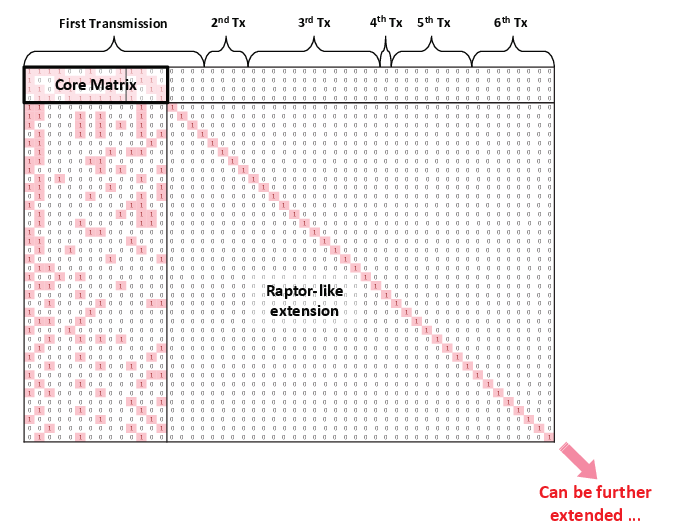}
    \caption{LDPC-based IR-HARQ (5G Base Graph 2)}
    \label{fig:LDPC-IR-HARQ}
\end{figure}

By contrast, 5G NR adopts a protograph-based raptor-like LDPC construction~\cite{LDPC:PBRL}, as shown in Fig.~\ref{fig:LDPC-IR-HARQ}. Each NR LDPC base graph contains:
\begin{itemize}
  \item A relatively high-density core (functionally similar to the outer codes of a Raptor code);
  \item A raptor-like low-density extension part (analogous to the inner codes of a Raptor code) that can generate additional parity checks on demand.
\end{itemize}

The key feature is that the extension parity bits are formed as linear combinations of (pseudo-)randomly chosen information bits. This directly realizes the rateless property: the encoder can, in principle, keep producing new, linearly independent parity bits without redesigning the code.

This design has several important implications for IR-HARQ:
\begin{enumerate}
  \item \emph{Fine-grained rate adaptation}: The raptor-like extension enables 1-bit (or very small step) granularity in rate adaptation. A higher-rate codeword is obtained by puncturing a lower-rate ``mother'' codeword; this nesting is inherent to the parity-check matrix construction and does not require separate code designs for each target rate.
  \item \emph{Single mother code, rate-independent construction}: The LDPC code is effectively constructed once (via the base graph and its lifting) and is independent of the final transmitted length. This is a natural fit to IR-HARQ, where the code rate is not known a priori and adapts to the channel via feedback.
  \item \emph{Near-uniform performance across a wide rate range}: For well-designed base graphs (e.g., 5G NR BG1 and BG2), the same mother code delivers capacity-approaching performance over a broad range of effective code rates. That is, code rate flexibility does not significantly compromise performance. This is crucial for IR-HARQ, where decoding must remain close to optimal after each incremental redundancy transmission, not only at one or two predefined rates.
\end{enumerate}

Thanks to these properties, 5G's raptor-like LDPC codes can fully exploit the potential of rateless IR-HARQ.

\section{Are polar codes rateless?}\label{section:polar}
The quick answer is: conventional polar codes~\cite{Polar:Arikan} cannot fully support rateless IR-HARQ. The fundamental obstacle comes from the length-dependent nature of polar code construction. This makes it impossible, in general, to build a single ``mother'' polar code from which all higher-rate versions are its nested codewords.

\subsection{Length-dependent construction of polar codes}\label{section:polar:length-dependent}
A polar code~\cite{Polar:Arikan} of length $N$ and dimension $K$ is constructed by selecting the $K$ most reliable polarized subchannels to carry information bits; the remaining $N-K$ least reliable subchannels are frozen to known values. The reliability of these subchannels:
\begin{itemize}
  \item depends on the underlying physical channel, and
  \item changes with both the code length and the rate-matching operation (puncturing/shortening).
\end{itemize}

Puncturing a code bit makes its corresponding bit-channel have effectively zero capacity; shortening yields absolute reliability (perfect a priori knowledge) while carrying no actual information. Consequently, different rate/length configurations lead to different reliability orderings and hence different information/frozen sets.

Early work on rate-compatible polar codes~\cite{Polar:QUP,Polar:WangLiuShorten} used on-the-fly reliability calculation that depends on instantaneous CSI (e.g., SNR), which is impractical in real systems because perfect, real-time CSI is unavailable.

5G polar codes~\cite{5G} circumvent this by adopting an offline, channel-independent design:
\begin{itemize}
  \item A fixed reliability sequence of length 1024 is precomputed;
  \item Rate matching (via carefully designed puncturing and shortening patterns) attempts to preserve this pre-defined reliability order for different lengths/rates.
\end{itemize}

This provides practical rate and length flexibility, but only at the level of designing a family of codes for different code rates - not a single mother code that is rate-compatible in the strict, rateless sense.

\subsection{Rate/length flexibility $\neq$ rateless IR-HARQ}\label{section:polar:notRateless}
Rateless IR-HARQ requires that:
\begin{itemize}
  \item A single mother code of length $N_{\max}$ and dimension $K$ is defined.
  \item In each transmission, a subset of the $N_{\max}$ coded bits are sent.
\end{itemize}

For 5G polar codes, this nesting property does not hold. In practice:
\begin{itemize}
  \item High-rate codes employ shortening.
  \item Low-rate codes employ puncturing.
  \item The optimal puncturing and shortening patterns (and hence the optimal information sets) are not compatible for different rates.
\end{itemize}

Thus, as the effective code rate decreases through incremental redundancy transmissions, the higher-rate code is not simply a subset of a lower-rate code defined on the same underlying polar transform. This violates a key structural requirement for rateless IR-HARQ.

\subsection{Existing polar IR-HARQ schemes}\label{section:polar:existingIR-HARQ}
There are two important works on polar IR-HARQ, one based on incremental freezing \cite{Polar:IF-HARQ} and the other is based on polarization matrix extension \cite{Polar:IR-HARQ}. For clarity, we first assume that there are two transmissions of length $N$ each.
\subsubsection{Incremental freezing}
\begin{figure}
    \centering
    \includegraphics[width=\linewidth]{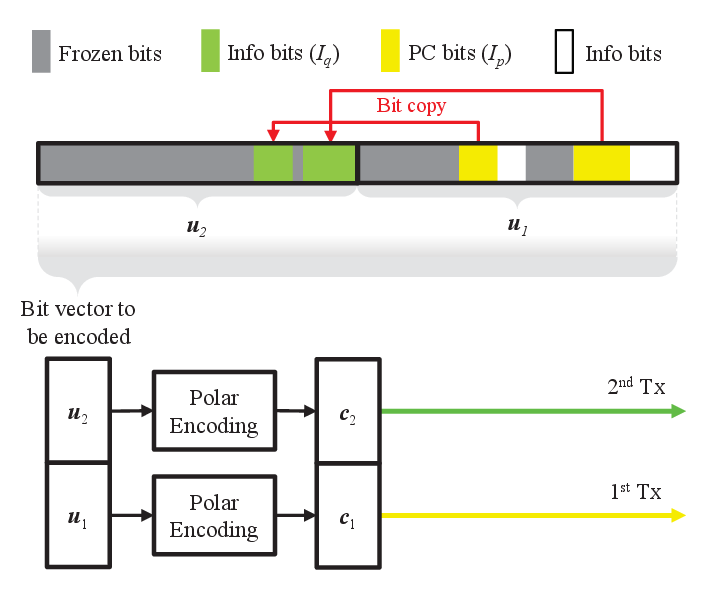}
    \caption{Polar-based IR-HARQ via incremental freezing}
    \label{fig:Polar-IF-HARQ}
\end{figure}

The incremental freezing scheme is illustrated in Fig.~\ref{fig:Polar-IF-HARQ} and works as follows:
\begin{itemize}
  \item In the first transmission, a length-$N$ polar code of rate $R_1 = K/N$ is used.
  \item In the second transmission, some unreliable bits from the first transmission are re-encoded into a lower-rate polar code and retransmitted.
  \item If they are successfully decoded, their values are regarded as known and frozen in a second decoding attempt of the first codeword - effectively reducing the its code rate.
\end{itemize}

This mechanism does exploit HARQ-type gains, but both transmissions are encoded and decoded as independent length-$N$ polar codes. They are not jointly decoded as a single length-$2N$ polar code. As a result, the scheme cannot harvest the full coding gain that would be available from a truly length-$2N$ polar code.

\subsubsection{Polarizing matrix extension}
To approach the performance of longer codes, \cite{Polar:IR-HARQ} proposes matrix extension:
\begin{itemize}
  \item The length-$N$ polar transform used in the first transmission is extended to form a larger polar transform (e.g., to length $2N$).
  \item After a retransmission, the two length-$N$ blocks are combined and jointly decoded as a length-$2N$ polar code.
  \item This can, in principle, yield coding and diversity gains closer to those of an actual length-$2N$ polar code.
\end{itemize}

\begin{figure}
    \centering
    \includegraphics[width=\linewidth]{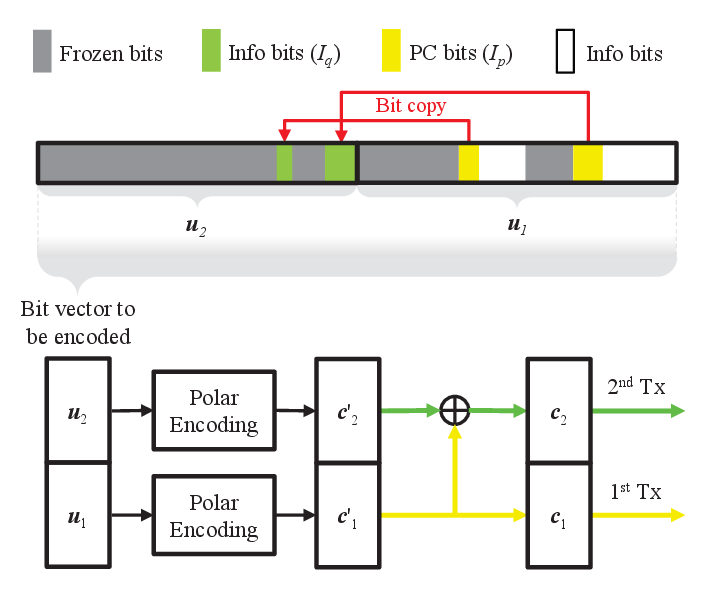}
    \caption{Polar-based IR-HARQ via polarizing matrix extension}
    \label{fig:Polar-IR-HARQ}
\end{figure}

The scheme is illustrated in Fig.~\ref{fig:Polar-IR-HARQ}. Specifically, since polar code construction is length-dependent, we let:
\begin{itemize}
  \item $\mathcal{I}_1$ be the optimal information set (size $K$) for length $N$,
  \item $\mathcal{I}_2$ be the optimal information set (size $K$) for length $2N$.
\end{itemize}

Typically, $\mathcal{I}_1 \neq \mathcal{I}_2$, which implies:
\begin{itemize}
  \item Some positions $\mathcal{I}_p \subset \mathcal{I}_1$ that are reliable at length $N$ become less reliable at length $2N$,
  \item Some new positions $\mathcal{I}_q = \mathcal{I}_2 \setminus \mathcal{I}_1$ become more reliable at length $2N$.
\end{itemize}

To retain optimality for both code length $N$ and $2N$, one approach is to map the same information bits to positions in $\mathcal{I}_p$ and $\mathcal{I}_q$, so that:
\begin{itemize}
  \item In the first transmission (length $N$), $\mathcal{I}_1$ is used as the information set, giving the optimal length-$N$ performance.
  \item In the combined length-$2N$ code, $\mathcal{I}_2$ is used, and the bits in $\mathcal{I}_p$ can be treated as parity-check frozen bits, since their values have already been recovered from $\mathcal{I}_q$.
\end{itemize}

This scheme improves over \cite{Polar:IF-HARQ} by more closely matching a longer polar code. However, it still does not provide a fully flexible rateless solution in general.

\subsection{Fundamental limitations}\label{section:polar:limitation}
For potential applications in 6G and beyond~\cite{6G:Horizon}, arbitrary transmissions length should be supported, and the inherent length-dependence of polar code construction leads to two critical limitations.

\subsubsection{No universal mother code}
For a truly rateless scheme, one desires a single mother code of length $N_{\max}$ and dimension $K$. For polar codes, this is unattainable in general because:
\begin{itemize}
  \item The optimal information set is tied to the actual code length and the rate matching pattern.
  \item Polar codes require re-optimizing the information/frozen set whenever the effective code length/rate changes.
\end{itemize}

Thus, there is no universal polar mother code with a rate-independent construction that supports arbitrary incremental redundancy in a strictly nested fashion.

\subsubsection{Loss of coding gain and flexibility for arbitrary retransmission lengths}
Even with matrix extension \cite{Polar:IR-HARQ}, optimal coding gain can often be guaranteed only for specific retransmission lengths (e.g., two equal-length transmissions). For arbitrary retransmission lengths, serious issues arise.

Consider an extreme example:
\begin{itemize}
  \item First transmission length: $E_1 = N$.
  \item Second transmission length: $E_2 = 1$.
\end{itemize}

To embed this into a length-$E^{(2)}=E_1 + E_2 = N+1$ polar transform via matrix extension, the extended part has $N-1$ punctured bits - they correspond to $N-1$ zero-capacity bit channels. The resulting polarized subchannels for the extended part are so unreliable that:
\begin{itemize}
  \item All the polarized subchannels associated with the extended part must be frozen.
  \item These frozen bit positions cannot later be converted into information bit positions in further retransmissions.
\end{itemize}

As such, any subsequent retransmissions will have no coding gain at all. Recall the basic $2 \times 2$ polar transform:
\begin{equation*}\label{polar_transform}
  (u_1, u_2) \mapsto (c_1, c_2) = (u_1 \oplus u_2, u_2).
\end{equation*}
If the upper-left input bit $u_1$ is frozen, then $c_1$ becomes a repetition of $u_2$. Now imagine $N$ such transforms in parallel. When all extended subchannels are frozen due to extreme puncturing, the new code bits corresponding to the retransmission are effectively repetitions of existing code bits from the initial transmission. They therefore contribute almost no additional coding gain; they mainly provide repetition gain rather than genuine new redundancy.

This phenomenon is not limited to the pathological case $E_2 = 1$; it reveals that:
\begin{itemize}
  \item Once a bit is frozen due to length-dependent construction under severe puncturing/shortening, it cannot be ``reclaimed'' for carrying new information in subsequent transmissions.
  \item Consequently, for many arbitrary IR patterns (variable $E_1, E_2, E_3,\dots$), the achievable coding gain is significantly constrained, and in some cases, additional transmissions bring little or no coding benefit.
\end{itemize}

\begin{remark}
Therefore, conventional polar codes - whether in their 5G form or in existing IR-HARQ schemes - are not rateless. Achieving genuine rateless IR-HARQ with polar codes would require fundamentally new constructions that decouple code design from blocklength in a way that preserves nesting and maintains near-optimal performance at all cumulative lengths.
\end{remark}

Based on the discussions in Section~\ref{section:intro}, \ref{section:existing3GPP} and \ref{section:polar}, we conclude that LDPC codes are rateless, but turbo codes and polar codes do not fully qualify as rateless, as summarized in Table~\ref{tab:codes-comp}. The goal of this work is to provide the rateless property for polar codes.

\begin{table}
    \centering
    \begin{tabular}{|c|c|c|c|c|}
    \hline
        Code Property & RM & Turbo & LDPC & Polar \\ \hline
        Fixed ($N,K$) & $\checkmark$ & $\checkmark$ & $\checkmark$ & $\checkmark$ \\ \hline
        Flexible ($N,K$) & $\times$ & $\checkmark$ 3G/4G & $\checkmark$ 5G & $\checkmark$ 5G \\ \hline
        Rateless (nested $N$) & $\times$ & $\halfcheck$ 3G/4G & $\checkmark$ 5G & This work \\ \hline
    \end{tabular}
    \caption{Code property of 3GPP standardized codes}
    \label{tab:codes-comp}
\end{table}

\section{Polar codes: channel-dependent or channel-independent?}\label{section:polar-question}
In this section, we revisit a fundamental dilemma that traces back to the very definition of polar codes, and explain why this original definition seems to prevent them from being channel-adaptive and thus inherently rateless. The key issue is the apparent channel dependence of polar code construction in theory, versus the channel-independent constructions that are widely used in practice.

\subsection{Channel-dependent in theory}\label{section:polar-question:channel-dependent}
In Ar{\i}kan's seminal paper~\cite{Polar:Arikan}, a polar code of length $N$ and dimension $K$ is defined by specifying the information set $\mathcal{I} \subset \{1,\dots,N\}$ of size $|\mathcal{I}| = K$, such that the polarized subchannels indexed by $\mathcal{I}$ are the most reliable ones. Formally, $\mathcal{I}$ is chosen to satisfy
\begin{equation}
Z(W_N^i) \leq Z(W_N^j),\quad \forall i \in \mathcal{I},\; j \in \bar{\mathcal{I}},
\end{equation}
where $W_N^i$ is the $i$-th polarized subchannel induced by the physical channel $W$, and $Z(\cdot)$ denotes the Bhattacharyya parameter (a reliability metric). Since the set $\{W_N^i\}$ and their reliabilities $Z(W_N^i)$ depend explicitly on the underlying physical channel $W$, Ar{\i}kan remarked that ``\emph{Polar codes are channel-specific designs: a polar code for one channel may not be a polar code for another}''~\cite{Polar:Arikan}.

Early polar code construction schemes fully embraced this channel dependence. Classical works such as \cite{Polar:DE1,Polar:DE2}, and \cite{Polar:GA} evaluate the reliability of the polarized subchannels with respect to a given channel $W$ and its SNR (or equivalent parameterization):
\begin{itemize}
  \item Density evolution (DE)-based methods \cite{Polar:DE1,Polar:DE2} track the full probability density functions of log-likelihood ratios (LLRs) through the polarization transform to obtain highly accurate reliability estimates.
  \item Gaussian approximation (GA)-based methods \cite{Polar:GA} approximate the LLR distributions as Gaussian and propagate their means/variances, trading off some accuracy for much lower complexity.
\end{itemize}
In all these cases, the construction explicitly requires a channel model and a channel parameter (e.g., SNR), and the resulting information set $\mathcal{I}$ is, by definition, channel-dependent.

\subsection{Channel-independent in practice}\label{section:polar-question:channel-independent}
During 5G NR standardization (around 2015), it became clear that strict channel-dependent constructions pose serious implementation challenges:
\begin{itemize}
  \item Lack of perfect channel knowledge: Perfect and static channel state information is unavailable at the transmitter in practical wireless systems, and even the receiver only has imperfect, time-varying estimates.
  \item Real-time complexity constraints: Even if perfect channel parameters were available, performing DE or GA online for each transport block or SNR operating point is prohibitively complex for hardware implementations, especially under stringent latency and power constraints.
\end{itemize}

To overcome these issues, several channel-independent polar code construction methods were proposed and evaluated. A representative approach is the Polarization Weight (PW) method \cite{Polar:PW}, which assigns to each polarized subchannel a reliability weight derived from its index using a simple, channel-independent ``beta-expansion'' formula~\cite{Polar:betaExpansion}. In 5G NR, the final solution is effectively a fixed global reliability ordering:
\begin{itemize}
  \item A universal reliability sequence of length 1024 is specified in the standard, obtained offline from extensive simulations, analytical approximations, and machine-learning-based optimizations~\cite{Polar:AIcoding}.
  \item For a given code length $N$ and dimension $K$, the information set $\mathcal{I}$ is chosen as the indices corresponding to the $K$ most reliable positions in this pre-defined sequence.
\end{itemize}

This design is channel-independent at run time: the transmitter and receiver simply use a look-up table. No channel model, no SNR parameter, and no on-the-fly DE/GA computation are required. This guarantees interoperability and keeps complexity extremely low.

Why do channel-independent constructions work in practice? The key observation is that practical systems operate near a specific ``working region'' of SNR and rate, not across all possible channel conditions. Consider a wireless system that adapts its modulation and coding scheme (MCS) so that the code rate $R = K/N$ roughly matches the available channel capacity. For example, a polar code of rate $R = 1/2$ is not designed to operate extremely high SNR (e.g., 20 dB) or extremely low SNR (e.g., -20 dB). Instead, it has an intended operating SNR region, say around 0$\sim$3 dB. Within such a working region:
\begin{itemize}
  \item The separation between ``good'' and ``bad'' subchannels, i.e., between $\mathcal{I}$ and $\bar{\mathcal{I}}$, becomes relatively stable and ``deterministic''.
  \item The reliability ordering within $\mathcal{I}$ (or within $\bar{\mathcal{I}}$) is irrelevant since polar codes are defined by $\mathcal{I}$ rather then the ordering within $\mathcal{I}$.
\end{itemize}

The above intuition can be turned into a conceptual procedure (Algorithm~\ref{alg:ci_reliability_order}) for deriving a universal reliability sequence from any given channel-dependent construction method.

\begin{algorithm}
\caption{Channel-independent reliability ordering}
\label{alg:ci_reliability_order}
\begin{algorithmic}[1]
\STATE \textbf{Input:} Number of subchannels $N$
\STATE \textbf{Output: }Reliability ordered sequence $Q$
\STATE $Q \gets \left[\ \right]$ \COMMENT{Initialize empty sequence}
\FOR{$k \gets 1$ \textbf{to} $N$}
    \STATE $R \gets k/N$ \COMMENT{Target code rate}
    \STATE $\mathit{SNR} \gets \mathcal{S}(R)$ \COMMENT{Operating SNR for rate $R$}
    \STATE $O \gets f(N, \mathit{SNR})$ \COMMENT{Channel-dependent reliability order}
    \STATE $\mathcal{I} \gets O[1:k]$ \COMMENT{Top $k$ reliable positions}
    \STATE $i \gets \mathcal{I} \setminus Q$ \COMMENT{Unique element in $\mathcal{I}$ not in $Q$}
    \STATE $Q \gets \left[\, i,\ Q\, \right]$ \COMMENT{Prepend new position}
\ENDFOR
\RETURN $Q$
\end{algorithmic}
\end{algorithm}

\subsection{The dilemma}\label{section:polar-question:dilemma}
While the channel-dependent nature of polar codes prevents low-complexity offline construction, the channel-independent approach also fails to support nested codewords with a fixed reliability ordering. Resolving this fundamental dilemma is essential for achieving rateless IR-HARQ.

\section{New polarization and decoding scheduling}\label{section:scheduling}
In this section, we introduce a more general polarization framework enabled by an additional degree of freedom: the \emph{decoding schedule}. Unlike the conventional setting, where the decoding order in successive-cancellation (SC) decoding is \emph{fixed}, we explicitly regard the decoding schedule as a design parameter. This new framework allows \emph{code-length adaptation} to be implemented at the decoder rather than the encoder, so that a single, \emph{fixed} code construction can achieve near-optimal performance over a wide range of effective code lengths. This framework paves the way toward \emph{rateless} polar coding schemes.

Classical channel polarization, starting from Ar{\i}kan's seminal work and followed by most subsequent studies, assumes a fixed ``$1 \rightarrow N$'' \emph{sequential} SC decoding schedule. Under this conventional paradigm, the reliability of the $i$-th synthesized subchannel is defined under the following implicit assumptions:
\begin{enumerate}
\item The SC decoding tree is traversed in a depth-first manner, typically from the leftmost leaf to the rightmost leaf.
\item When making a hard decision on the $i$-th bit, the previously decoded bits $1,2,\dots,i-1$ are assumed to have been decoded correctly.
\end{enumerate}

These assumptions are deeply embedded in the standard analysis of polarization, including the recursive evolution of subchannel capacity and reliabilities. If either of the above assumptions is relaxed, the evaluation methodology for polarization and the resulting subchannel reliabilities can change substantially.

\subsection{Generalized single-step polar transform}
Recall that in Ar{\i}kan's seminal work, the Bhattacharyya parameter $Z(W)$ is used to measure the reliability of a binary-input discrete memoryless channel (B-DMC) $W$, which provides an upper bound on the probability of a maximum-likelihood (ML) decision error.

For the standard single-step polarization transform $(W, W) \to (W', W'')$ under conventional decoding order ($W'$ followed by $W''$), the recursive relations are given by:
\begin{align}
    Z(W') &\leq 2Z(W) - Z(W)^2 \label{eq:Z_hom_minus}\\
    Z(W'') &= Z(W)^2 \label{eq:Z_hom_plus}
\end{align}
where equalities hold if and only if $W$ is a Binary Erasure Channel (BEC).

In the more general case where the input channels $W_1$ and $W_2$ are not statistically identical, the generalized single-step transform $(W_1, W_2) \to (W', W'')$ yields:
\begin{align}
    Z(W') &\leq Z(W_1) + Z(W_2) - Z(W_1)Z(W_2) \label{eq:Z_het_minus}  \\
    Z(W'') &= Z(W_1)Z(W_2) \label{eq:Z_het_plus}
\end{align}
Consistent with the homogeneous construction, the equality for $Z(W')$ hold if and only if both $W_1$ and $W_2$ are BECs.

Throughout this paper, we assume a BEC in the recursive calculation of Bhattacharyya parameters. For Additive White Gaussian Noise (AWGN) channels, the Bhattacharyya parameter calculated under the BEC assumption serves as an upper bound. This generalized recursion provides the analytical tool for evaluating subchannel reliability under arbitrary decoding schedules.

\subsection{Code-length adaptation via decoding scheduling}
\subsubsection{An $N=2$ example}
\begin{figure}
    \centering
    \includegraphics[width=0.6\linewidth]{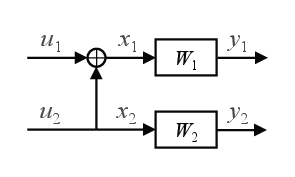}
    \caption{Generalized single-step polar transform}
    \label{fig:Polar-2x2}
\end{figure}

We first consider the simplest generalized single-step transform $(W_1, W_2) \to (W', W'')$, shown in Fig.~\ref{fig:Polar-2x2}. Let $u_1$ and $u_2$ denote the information bits. These are mapped to code bits $x_1 = u_1 \oplus u_2$ and $x_2 = u_2$, which are then transmitted over channels $W_1$ and $W_2$, resulting in the observations $y_1$ and $y_2$, respectively.

There can be two decoding orders:
\begin{itemize}
    \item \textbf{Scheduling $u_1 \to u_2$:}
    This is the conventional decoding order, and the standard recursive relations \eqref{eq:Z_het_minus}, \eqref{eq:Z_het_plus} hold.

    \item \textbf{Scheduling $u_2 \to u_1$:}
    If we invert the schedule, we effectively treat $u_2$ as the first bit to be decoded. In this case, the transform effectively reduces to $W_2 \to W''$. Once $u_2$ is decided, $u_1 = x_1 \oplus u_2$ effectively reduces to $W_1 \to W'$.
\end{itemize}

Under the reverse decoding order ($W''$ followed by $W'$), the recursive relations become:
\begin{align}
    Z(W') &= Z(W_1) \label{eq:Z_het21_minus}  \\
    Z(W'') &= Z(W_2) \label{eq:Z_het21_plus}
\end{align}

\subsubsection{An $(N\geq5, K=4)$ rateless polar coding example}\label{section:scheduling:length-adapt:K4N5678}
To demonstrate the principle of code-length adaptation through decoding scheduling, we examine a rateless polar coding scenario with an information length of $K=4$. In this framework, the code length $N \in \{5, 6, 7, 8, \dots\}$ is not fixed during construction or encoding, requiring the decoder to adapt to varying degrees of puncturing.

For polar codes, a set of partial orders exists that reveals deterministic reliability relationships applicable to any binary-input memoryless symmetric channel (BMSC)~\cite{Polar:PO-RM}. For $N=\{5,6,7,8\}$ under sequential puncturing, the subchannel reliability ordering is always:
\begin{align*}
&Z(W_N^1)\geq Z(W_N^2)\geq Z(W_N^3)\geq Z(W_N^5)\\
&\geq Z(W_N^4)\geq Z(W_N^6)\geq Z(W_N^7)\geq Z(W_N^8).
\end{align*}
To optimize performance for $K=4$, we allocate the information bits $\mathbf{u} = \{u_1, u_2, u_3, u_4\}$ to the most reliable indices identified by the information set $\mathcal{I} = \{4, 6, 7, 8\}$.

\begin{figure}
    \centering
    \includegraphics[width=\linewidth]{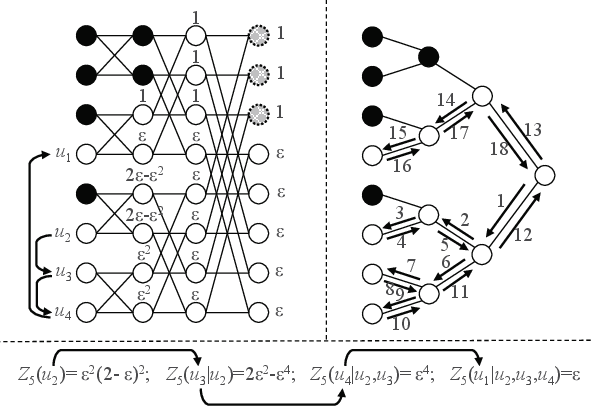}
    \caption{An example of the $u_2 \to u_3 \to u_4 \to u_1$ schedule and its recursive evolutions.}
    \label{fig:Fig-Polar-N5K4}
\end{figure}

We analyze two candidate decoding schedules:
\begin{itemize}
    \item \textbf{Schedule 1 ($S_1$):} $u_1 \to u_2 \to u_3 \to u_4$. This is the standard schedule. Its recursive evolution follows \eqref{eq:Z_het_minus}, \eqref{eq:Z_het_plus}.
    \item \textbf{Schedule 2 ($S_2$):} $u_2 \to u_3 \to u_4 \to u_1$. An example of this schedule is illustrated in Fig.~\ref{fig:Fig-Polar-N5K4} for $N=5,K=4$. Its recursive evolutions follow both \eqref{eq:Z_het_minus}, \eqref{eq:Z_het_plus} and \eqref{eq:Z_het21_minus}, \eqref{eq:Z_het21_plus}, also shown in Fig.~\ref{fig:Fig-Polar-N5K4}.
\end{itemize}
The performance is evaluated using the Bhattacharyya parameter $Z(W)$ to bound the block error probability $P_B \approx 1 - \prod_{i=1}^K (1 - Z(u_i|\dots))$.

\paragraph{Case 1: Code Length $N=5$}
With the first three code bits punctured, the channel reliabilities are $Z(x_1)=Z(x_2)=Z(x_3)=1$ and $Z(x_4)=\dots=Z(x_8)=\varepsilon$.

\begin{itemize}
    \item \textbf{Schedule $S_1$:} Applying the recursive relations yields:
    \begin{align*}
        Z_5(u_1) &= 2\varepsilon - \varepsilon^2 \\
        Z_5(u_2|u_1) &= \varepsilon^2(2 - \varepsilon)(1 + \varepsilon - \varepsilon^2) \\
        Z_5(u_3|u_1, u_2) &= \varepsilon^2 + \varepsilon^3 - \varepsilon^5 \\
        Z_5(u_4|u_1, u_2, u_3) &= \varepsilon^5
    \end{align*}

    \item \textbf{Schedule $S_2$:} Reordering the decoding sequence results in:
    \begin{align*}
        Z_5(u_2) &= \varepsilon^2(2 - \varepsilon)^2 \\
        Z_5(u_3|u_2) &= 2\varepsilon^2 - \varepsilon^4 \\
        Z_5(u_4|u_2, u_3) &= \varepsilon^4 \\
        Z_5(u_1|u_2, u_3, u_4) &= \varepsilon
    \end{align*}

    \item \textbf{Comparison:} For $N=5$, $S_2$ is strictly superior for all $\varepsilon \in (0, 1)$.
    \begin{proof}
    Let $\Delta Z_5 = P_B(S_1) - P_B(S_2)$. Algebraic simplification yields:
    \begin{align*}
        &\Delta Z_5 =\\
        & -\varepsilon(\varepsilon - 1)^6(\varepsilon + 1) \cdot \\
        &(\varepsilon^9 - 2\varepsilon^7 - 4\varepsilon^6 - 2\varepsilon^5 - 2\varepsilon^4 - 3\varepsilon^3 - 2\varepsilon^2 - \varepsilon - 1) \\
        =& (\text{Negative}) \times (\text{Positive}) \times (\text{Positive}) \times (\text{Negative})\\
        >& 0.
    \end{align*}
    Thus, $S_1$ consistently yields a higher error probability than $S_2$ at this code length.
    \end{proof}
\end{itemize}

\paragraph{Case 2: Code Length $N=6$}
With $x_1, x_2$ punctured, the input reliabilities are $Z(x_1)=Z(x_2)=1$ and $Z(x_3)=\dots=Z(x_8)=\varepsilon$.

\begin{itemize}
    \item \textbf{Schedule $S_1$:}
    \begin{align*}
        Z_6(u_1) &= (2\varepsilon - \varepsilon^2)^2\\
        Z_6(u_2|u_1) &= (\varepsilon + \varepsilon^2 - \varepsilon^3)^2 \\
        Z_6(u_3|u_1, u_2) &= 2\varepsilon^3 - \varepsilon^6\\
        Z_6(u_4|u_1, u_2, u_3) &= \varepsilon^6
    \end{align*}

    \item \textbf{Schedule $S_2$:}
    \begin{align*}
        Z_6(u_2) &= \varepsilon^2(2 - \varepsilon)^2\\
        Z_6(u_3|u_2) &= 2\varepsilon^2 - \varepsilon^4 \\
        Z_6(u_4|u_2, u_3) &= \varepsilon^4\\
        Z_6(u_1|u_2, u_3, u_4) &= \varepsilon^2
    \end{align*}

    \item \textbf{Comparison:} At $N=6$, the optimal schedule becomes channel-dependent.
    \begin{proof}
    \begin{align*}
        &\Delta Z_6 = P_B(S_1) - P_B(S_2)=\\
        & \varepsilon^2(\varepsilon - 1)^6(\varepsilon + 1)^2(\varepsilon^2 - 2\varepsilon - 1)\cdot\\
            &(\varepsilon^{10} - 4\varepsilon^7 - 2\varepsilon^6 - 4\varepsilon^5 - 2\varepsilon^3 + \varepsilon^2 + 2)\\
            =&(\text{Positive}) \times (\text{Positive})\times(\text{Positive}) \times (\text{Negative})\\
            &\times f(\varepsilon),
    \end{align*}
    The difference $\Delta Z_6$ is governed by the function $f(\varepsilon)$.
    Since $f(0)=2$ and $f(1)=-8$, there exists a unique root $\varepsilon_{th} \approx 0.746$. Consequently:
    \begin{itemize}
        \item $S_1$ is optimal for $0 < \varepsilon < 0.746$.
        \item $S_2$ is optimal for $0.746 < \varepsilon < 1$.
    \end{itemize}
    \end{proof}
\end{itemize}

Detailed proofs for $N \in \{7, 8\}$ are provided in Appendix~\ref{section:appendix:channel-dependent-scheduling-proof}, which similarly demonstrate channel-dependent switching at $\varepsilon_{th} \approx 0.920$ and $0.965$, respectively.

\begin{remark}
This behavior indicates that polar codes exhibit channel-dependency not only in their construction (frozen set selection) but also in their decoding scheduling.
\end{remark}

For practical operating regions ($\varepsilon < 0.5$), the optimal scheduling is primarily code-length-dependent, as shown in Table \ref{tab:decoding_schedule}. This dependency arises because different code lengths correspond to distinct puncturing patterns and varying distributions of zero-capacity input channels.
\begin{table}[ht]
\centering
\caption{Optimal Decoding Schedule for $K=4$}
\label{tab:decoding_schedule}
\begin{tabular}{|c|c|c|c|c|}
\hline
Decoding Schedule & \textbf{N=5} & \textbf{N=6} & \textbf{N=7} & \textbf{N=8} \\
\hline
$u_1 \to u_2 \to u_3 \to u_4$ & $\times$ & $\checkmark$ & $\checkmark$ & $\checkmark$ \\
\hline
$u_2 \to u_3 \to u_4 \to u_1$ & $\checkmark$ & $\times$ & $\times$ & $\times$ \\
\hline
\end{tabular}
\end{table}

The practical implication is that by employing channel-aware decoding scheduling, we can maintain a \textit{fixed code construction} while adapting to fluctuating channel conditions or code lengths to ensure optimal performance.

\begin{figure}
    \centering
    \includegraphics[width=\linewidth]{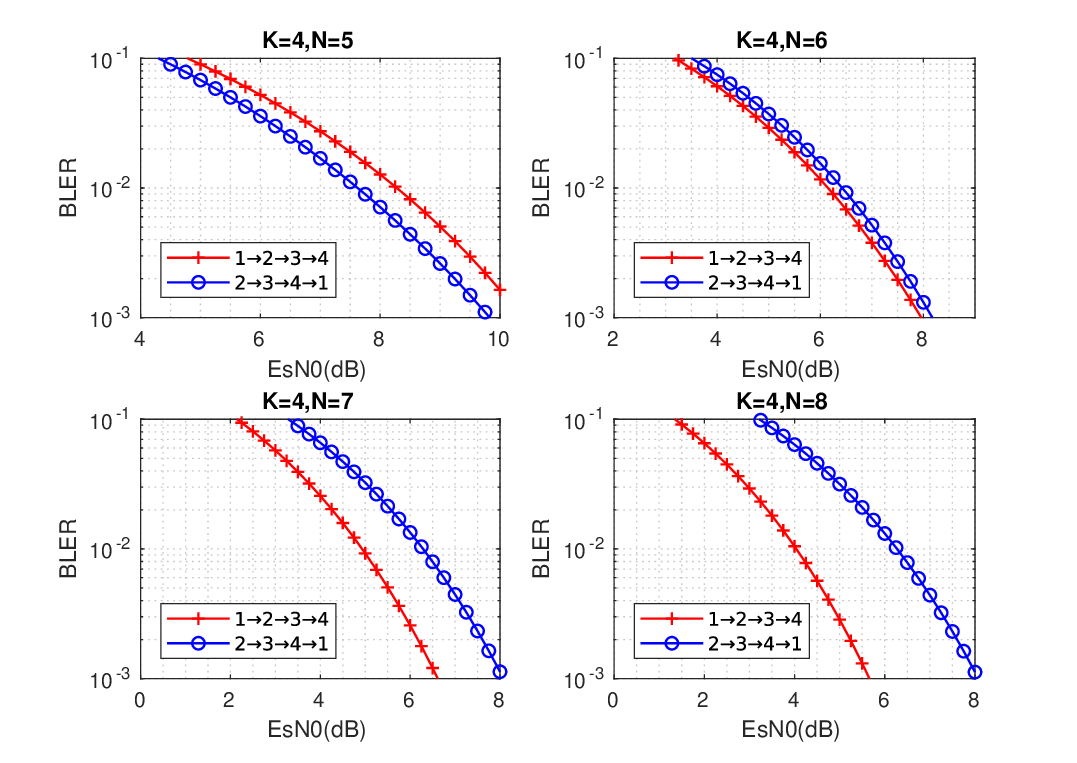}
    \caption{Performance comparison of different decoding scheduling for $(N\geq5, K=4)$.}
    \label{fig:Fig-Polar-K4N5678-BLER}
\end{figure}

Although the analysis of schedules $S_1$ and $S_2$ assumes a BEC, the code-length-dependent scheduling phenomenon is observed in AWGN channels as well, as demonstrated by the following simulation results. Fig.~\ref{fig:Fig-Polar-K4N5678-BLER} presents block error rate (BLER) results for $K=4, N=5,6,7,8$, which clearly demonstrates that the \emph{decoding order is a critical design parameter}. In traditional polar codes, the decoding order is fixed. However, by enabling flexible decoding scheduling, we can achieve code-length adaptation.

\subsubsection{SC decoding with arbitrary scheduling}
The fundamental shift from $1\to N$ sequential SC is the introduction of a decoding schedule $\mathcal{S}$.
\begin{quote}
\textbf{Definition:} A decoding schedule $\mathcal{S} = \{s_1, s_2, \dots, s_K\}$ is an ordered permutation of the information indices in $\mathcal{I}$ that dictates the sequence in which leaf nodes are visited and hard-decided.
\end{quote}

The decoding schedule $\mathcal{S}$ introduces a degree of freedom to adapt to different channel conditions or code lengths, such as punctured codes, by prioritizing the decoding of subtrees with higher available mutual information. If the schedule dictates that $u_j$ is decoded before $u_i$ (where $j > i$), the decoder prioritizes the lower subtree of the corresponding polar transform.

The decoder operates on LLRs. For any bit $x$, the LLR is defined as:
\begin{equation}
\lambda(x) = \ln \frac{P(y|x=0)}{P(y|x=1)}
\end{equation}
where $y$ is the channel observation. We denote $\boldsymbol{\lambda}_d$ as the LLR vector at depth $d$ of the decoding tree, and $\boldsymbol{\beta}_d$ as the corresponding vector of hard-decision partial sums.

We define the three primary processing functions:
\begin{itemize}
  \item $f$-function: Soft estimate of $u_1$ when $u_2$ is unknown.
        \begin{align}
            \lambda_{u_1} &= f(\lambda_{x_1}, \lambda_{x_2})\nonumber \\
            &= 2 \operatorname{tanh}^{-1} \left( \operatorname{tanh} \frac{\lambda_{x_1}}{2} \cdot \operatorname{tanh} \frac{\lambda_{x_2}}{2} \right)
        \end{align}
  \item $g$-function: Soft estimate of $u_2$ given $\hat{u}_1$.
        \begin{align}
            \lambda_{u_2} = g(\lambda_{x_1}, \lambda_{x_2}, \hat{u}_1) = \lambda_{x_2} + (-1)^{\hat{u}_1} \lambda_{x_1}
        \end{align}
  \item $h$-function (newly introduced reverse cancellation): Soft estimate of $u_1$ given $\hat{u}_2$.
        \begin{align}
            \lambda_{u_1} = h(\lambda_{x_1}, \hat{u}_2) = (-1)^{\hat{u}_2} \lambda_{x_1}
        \end{align}
\end{itemize}
The whole decoding process is decomposed into these basic transforms, which are recursively processed according to the sequence defined in $\mathcal{S}$.

To handle arbitrary decoding orders while maintaining memory efficiency, we propose a generalized SC decoding framework in Algorithm~\ref{alg:generalized_sc}.

\begin{algorithm}
\caption{Successive Cancellation Decoding with Arbitrary Schedule}
\label{alg:generalized_sc}
\begin{algorithmic}[1]
\STATE \textbf{Input:} Channel LLRs $\boldsymbol{\lambda}_{ch}$, Info set $\mathcal{I}$, Schedule $\mathcal{S} = \{s_1, \dots, s_K\}$
\STATE \textbf{Output:} Decoded vector $\hat{\mathbf{u}}$

\STATE $\hat{\mathbf{u}} \gets \text{vector of size } N \text{ initialized to } \text{None}$
\STATE $p \gets 1$ \COMMENT{Global schedule pointer}

\WHILE{$p \le K$}
    \IF{$\hat{\mathbf{u}}[s_p] \text{ is None}$} \label{alg:line:start}
        \STATE $\text{RecursiveDecode}(n, 1, \boldsymbol{\lambda}_{ch})$
    \ELSE
        \STATE $p \gets p + 1$
    \ENDIF
\ENDWHILE

\STATE \textbf{Function} \texttt{RecursiveDecode}($d, v, \boldsymbol{\lambda}$)
    \IF{$p > K \text{ OR } s_p \notin \mathcal{L}(d, v)$}
        \STATE \textbf{goto} line \ref{alg:line:start} \COMMENT{Exit subtree and return to root} \label{alg:line:break}
    \ENDIF

    \IF{$d = 0$}
        \STATE $\hat{\mathbf{u}}[v] \gets (v \in \mathcal{I} \text{ and } \boldsymbol{\lambda} < 0) ? 1 : 0$
        \STATE $p \gets p + 1$
        \RETURN $\hat{\mathbf{u}}[v]$
    \ENDIF

    \STATE $s_{target} \gets \mathcal{S}[p]$
    \STATE $(\boldsymbol{\lambda}_1, \boldsymbol{\lambda}_2) \gets \text{Split } \boldsymbol{\lambda} \text{ into upper/lower halves}$

    \IF{$s_{target} \in \mathcal{L}(d-1, 2v-1)$}
        \STATE \COMMENT{Standard Order: Upper subtree first}
        \STATE $\boldsymbol{\lambda}_{up} \gets f(\boldsymbol{\lambda}_1, \boldsymbol{\lambda}_2)$
        \STATE $\beta_1 \gets \texttt{RecursiveDecode}(d-1, 2v-1, \boldsymbol{\lambda}_{up})$
        \STATE $\boldsymbol{\lambda}_{low} \gets g(\boldsymbol{\lambda}_1, \boldsymbol{\lambda}_2, \beta_1)$
        \STATE $\beta_2 \gets \texttt{RecursiveDecode}(d-1, 2v, \boldsymbol{\lambda}_{low})$
    \ELSE
        \STATE \COMMENT{Reversed Order: Lower subtree first}
        \STATE $\boldsymbol{\lambda}_{low} \gets \boldsymbol{\lambda}_2$
        \STATE $\beta_2 \gets \texttt{RecursiveDecode}(d-1, 2v, \boldsymbol{\lambda}_{low})$
        \STATE $\boldsymbol{\lambda}_{up} \gets h(\boldsymbol{\lambda}_1, \beta_2)$
        \STATE $\beta_1 \gets \texttt{RecursiveDecode}(d-1, 2v-1, \boldsymbol{\lambda}_{up})$
    \ENDIF

    \RETURN $(\beta_1 \oplus \beta_2, \beta_2)$ \COMMENT{Propagate partial sums to parent}
\end{algorithmic}
\end{algorithm}

In this algorithm, we define $\mathcal{L}(d, v)$ as the set of leaf indices descendant from node $v$ at depth $d$. For a polar code of length $N=2^n$, this is defined as:
\begin{align*}
\mathcal{L}(d, v) = { i \in \mathbb{Z} : (v-1) 2^d + 1 \leq i \leq v 2^d }.
\end{align*}
When the next bit in the schedule $s_p$ falls outside the current subtree $\mathcal{L}(d, v)$, the algorithm executes a global jump (Line \ref{alg:line:break}) that immediately terminates the recursion stack and unwinds to the root (Line \ref{alg:line:start}). This ``break'' mechanism allows the decoder to re-enter the trellis for the new target index. Because the decision vector $\hat{\mathbf{u}}$ is persistent, previously computed hard decisions are preserved and utilized in subsequent decoding.

The proposed arbitrary scheduling framework can be naturally extended to Successive Cancellation List (SCL) decoding \cite{Polar:List_Tal,Polar:CA_List_Niu}. In the SCL variant, the global decision vector $\hat{\mathbf{u}}$ is replaced by a set of $L$ path candidates. When the \texttt{RecursiveDecode} function reaches a leaf node ($d=0$) in the information set $\mathcal{I}$, each surviving path splits into two (representing $u_i = 0$ and $u_i = 1$), and the $L$ most likely paths are retained based on their Path Metrics (PM).

\subsubsection{Channel-aware decoding scheduling}
Given the generalized SC framework that supports arbitrary decoding orders, the optimization problem of interest is to determine the decoding schedule $\mathcal{S}$ that minimizes the block error rate for a specific channel condition or puncturing pattern.

Formally, let $\mathcal{P}(\mathcal{I})$ denote the set of all $K!$ permutations of the information set $\mathcal{I}$. The optimal schedule $\mathcal{S}^*$ is defined as:
\begin{equation}
    \mathcal{S}^* = \arg \min_{\mathcal{S} \in \mathcal{P}(\mathcal{I})} P_e(\mathcal{S}).
\end{equation}
While an exhaustive search of $K!$ permutations is computationally prohibitive, we propose a sub-optimal yet computationally efficient greedy approach.

The proposed scheduling algorithm prioritizes the next bit to be decoded as the most critical one. This prioritization is necessitated by the inherent nature of sequential decoding: if the next bit is decoded incorrectly, the entire block is rendered erroneous. By addressing this bottleneck - ensuring the most reliable bit is decided first - the algorithm maximizes the probability of a successful decoding trajectory.

The reliability of each information subchannel is characterized by its bit error probability (or Bhattacharyya parameter) $Z$. Unlike conventional natural index ordered decoding, where the reliability of the $k$-th bit is evaluated assuming bits $u_1, \dots, u_{k-1}$ are known, our framework evaluates the error probability of $u_k$ conditioned on the set of decided bits $u_{s_1}, \dots, u_{s_{k-1}}$. The recursive density evolution defined in \eqref{eq:Z_het_minus}--\eqref{eq:Z_het21_plus} is used to track these reliabilities. The greedy scheduling procedure is formalized in Algorithm~\ref{alg:greedy_scheduling}.

\begin{algorithm}
\caption{Channel-Aware Decoding Scheduling}
\label{alg:greedy_scheduling}
\begin{algorithmic}[1]
\STATE \textbf{Input:} Information set $\mathcal{I}$, Channel condition $\boldsymbol{\epsilon}$
\STATE \textbf{Output:} Decoding schedule $\mathcal{S}$

\STATE $\mathcal{S} \gets ()$ \COMMENT{Initialize as an empty ordered sequence}
\STATE $\mathcal{U} \gets \mathcal{I}$ \COMMENT{Initialize set of unscheduled bits}

\WHILE{$\mathcal{U}$ is not empty}
    \STATE \COMMENT{Find the bit with the lowest conditional error probability}
    \STATE $s^* \gets \arg \min_{k \in \mathcal{U}} Z_N(u_k | \mathcal{S})$
    \STATE Append $s^*$ to $\mathcal{S}$
    \STATE $\mathcal{U} \gets \mathcal{U} \setminus \{s^*\}$
\ENDWHILE
\RETURN $\mathcal{S}$
\end{algorithmic}
\end{algorithm}

To demonstrate the efficacy of the greedy approach, we revisit the example with $K=4, N \in \{5, 6, 7, 8\}$, representing different puncturing patterns and channel conditions.
\begin{itemize}
    \item \textbf{Case $N=5$:} The unconditional error probability $Z_5(u_2) = \epsilon^2(2 - \epsilon)^2$ is the minimum among the set $\{u_1, u_2, u_3, u_4\}$. Thus, $u_2$ is scheduled first. Conditioned on $u_2$, the lowest reliability belongs to $u_3$ where $Z_5(u_3|u_2) = 2\epsilon^2 - \epsilon^4$. Finally, comparing the remaining bits, we find $Z_5(u_4 | u_2, u_3) = \epsilon^4$ is lower than that of $u_1$. The resulting optimal greedy schedule is $\mathcal{S} = \{u_2, u_3, u_4, u_1\}$.

    \item \textbf{Case $N=6$:} A tie occurs between $u_1$ and $u_2$ as $Z_6(u_1) = Z_6(u_2) = \epsilon^2(2 - \epsilon)^2$. To break the tie, we ``look ahead'' to evaluate the next bits $Z_6(u_2|u_1) = (\epsilon + \epsilon^2 - \epsilon^3)^2$ while $Z_6(u_3|u_2) = 2\epsilon^2 - \epsilon^4$. Since $Z_6(u_2|u_1) < Z_6(u_3|u_2)$ for all $\epsilon \in (0, 1)$, the algorithm selects $u_1$ followed by $u_2$. The complete schedule is determined as $\mathcal{S} = \{u_1, u_2, u_3, u_4\}$.

    \item \textbf{Case $N \in \{7, 8\}$:} For higher values of $N$, the channel provides sufficient redundancy for $u_1$ to become the most reliable bit, with $Z_7(u_1) = (2\epsilon - \epsilon^2)^3$ and $Z_8(u_1) = (2\epsilon - \epsilon^2)^4$. Both metrics are superior to $Z(u_2)$ in their respective cases, causing Algorithm~\ref{alg:greedy_scheduling} to converge to the standard sequential schedule $\mathcal{S} = \{u_1, u_2, u_3, u_4\}$.
\end{itemize}

These results corroborate the analytical finding in Section~\ref{section:scheduling:length-adapt:K4N5678} that for heavily punctured codes, the optimal decoding schedule may deviate from the natural index order.

Although Algorithm~\ref{alg:greedy_scheduling} assumes a BEC, its heuristic applies to other channels as well. For AWGN channels, the conditional error probability in Step 7 can be calculated through density evolution or Gaussian approximation. The simulation results in Section~\ref{section:application:simulation} demonstrate its efficacy in AWGN channels.

\subsubsection{New degree of freedom}
\begin{figure}
    \centering
    \includegraphics[width=\linewidth]{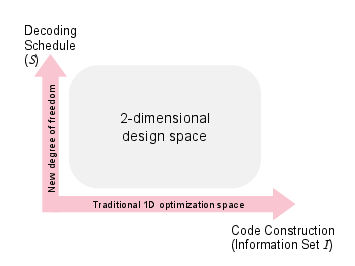}
    \caption{Expanding the Design Space: Code Construction ($\mathcal{I}$) $\times$ Decoding Schedule ($\mathcal{S}$)}
    \label{fig:new-DoF}
\end{figure}

The fundamental essence of this contribution lies in the introduction of a new degree of freedom within the polar code design space, as illustrated in Fig.~\ref{fig:new-DoF}.

Traditionally, for a given set of channel statistics or a specific rate-matching scheme, the optimization of polar codes was confined to code construction - specifically the selection of information and frozen sets. By formalizing the decoding schedule as an independent design variable, we transition from a one-dimensional optimization problem to a dual-degree-of-freedom framework.

\section{Flexible IR-HARQ in Wireless Communications}\label{section:application}
In this section, we apply the proposed arbitrary decoding schedule to the IR-HARQ problem. By synergistically combining a fixed nested code construction with dynamic scheduling, polar codes can seamlessly adapt to arbitrary block lengths.

\subsection{Nested Code Construction for Arbitrary Lengths}
To achieve ratelessness, the transmitted code length $E$ must be allowed to vary continuously between a minimum mother code length $N_{\min}$ and a maximum length $N_{\max}$. We achieve this by applying our decoding scheduling framework to a nested polar code construction.

The core mechanism of this construction is a ``one-to-many'' information bit mapping, where a single information bit is repeated (or copied) across multiple polarized subchannels in different subblocks. This introduces parity-check equations across the subblocks. The construction process is detailed in Algorithm~\ref{alg:nested_construction}.

\begin{algorithm}
\caption{Nested Polar Code Construction}
\label{alg:nested_construction}
\begin{algorithmic}[1]
\STATE \textbf{Input:} Minimum length $N_{\min}$, Maximum length $N_{\max}$, Information length $K$
\STATE \textbf{Output:} Global information set $\mathcal{I}_{\max}$, Mapping relations

\STATE $N \gets N_{\min}$
\STATE $\mathcal{I} \gets \emptyset$, $\mathcal{I}_{\max} \gets \emptyset$
\WHILE{$N \leq N_{\max}$}
\STATE Construct optimal information set $\mathcal{I}'$ for PolarCode($N, K$)
\STATE $\mathcal{I}_p \gets \mathcal{I} \setminus \mathcal{I}'$ \COMMENT{Bits unique to previous block}
\STATE $\mathcal{I}_q \gets \mathcal{I}' \setminus \mathcal{I}$ \COMMENT{New bits in current block}
\STATE $u_{\mathcal{I}_q} \gets u_{\mathcal{I}_p}$ \COMMENT{Establish bit-copy mapping (detailed in Sec. \ref{section:application:mapping})}
\STATE $\mathcal{I} \gets \mathcal{I}'$
\STATE $\mathcal{I}_{\max} \gets \mathcal{I}_{\max} \cup \mathcal{I}$
\STATE $N \gets N \times 2$
\ENDWHILE
\RETURN $\mathcal{I}_{\max}$
\end{algorithmic}
\end{algorithm}

With the encoding structure fixed, the remaining challenge is to design a decoding algorithm that optimizes performance for any arbitrary transmitted length $E$. Specifically, we must adapt the generalized SC decoding (Algorithm~\ref{alg:generalized_sc}) to leverage the parity-check relationships generated by the one-to-many bit mapping.

\subsection{Parity-Check Successive Cancellation Decoding with Arbitrary Schedule}
The ``bit copy'' operations in Algorithm~\ref{alg:nested_construction} mean that an information bit exists in multiple bit channels across different subblocks. This creates an explicit parity-check relationship: if a bit $u_i$ is copied to $u_j$, then $\hat{u}_i = \hat{u}_j$.

Consequently, if the schedule dictates that $u_i$ is decoded before $u_j$, the decision $\hat{u}_i$ instantly determines $u_j$. The bit $u_j$ effectively transitions from an unknown information bit to a known frozen bit for the remainder of the decoding process. We describe the adapted version of the generalized SC algorithm in a concise form (Algorithm~\ref{alg:generalized_sc_with_pc}), highlighting only the functional modifications while omitting unchanged components. Specifically, we introduce a tracking mechanism that updates all associated bit copies the moment a decision is made at the leaf node ($d=0$).

\begin{algorithm}
\caption{Parity-Check SC Decoding with Arbitrary Schedule (Concise)}
\label{alg:generalized_sc_with_pc}
\begin{algorithmic}[1]
\STATE \textbf{Input:} Channel LLRs $\boldsymbol{\lambda}_{ch}$, Info set $\mathcal{I}$, Schedule $\mathcal{S}$, Copy-sets $\mathcal{C}$
\STATE \textbf{Output:} Decoded vector $\hat{\mathbf{u}}$

\STATE \COMMENT{Initialization and main WHILE loop remain identical to Algorithm \ref{alg:generalized_sc}}
\STATE \dots

\STATE \textbf{Function} \texttt{RecursiveDecode}($d, v, \boldsymbol{\lambda}$)
\IF{$p > K \text{ OR } s_p \notin \mathcal{L}(d, v)$}
\STATE \textbf{goto} Main Loop \COMMENT{Exit subtree and return to root}
\ENDIF

\IF{$d = 0$}
    \STATE $\hat{u}_{temp} \gets (\boldsymbol{\lambda} < 0) ? 1 : 0$
    \FORALL{$w \in \text{CopySet}(v, \mathcal{C})$} \label{alg:line:pc_update}
        \STATE $\hat{\mathbf{u}}[w] \gets \hat{u}_{temp}$ \COMMENT{Instantly freeze all copied bits}
    \ENDFOR
    \STATE $p \gets p + 1$
    \RETURN $\hat{u}_{temp}$
\ENDIF

\STATE \COMMENT{Standard/Reversed recursive LLR calculations remain identical to Algorithm \ref{alg:generalized_sc}}
\STATE \dots

\RETURN $(\beta_1 \oplus \beta_2, \beta_2)$
\end{algorithmic}
\end{algorithm}

\subsection{Parity-Check Channel-Aware Decoding Scheduling}
The channel-aware scheduling algorithm is also adapted to account for the parity-check relationships defined by the copy-sets $\mathcal{C}$. This adaptation is formalized in Algorithm~\ref{alg:pc_greedy_scheduling}.

Once a bit $s^*$ is selected by the scheduler, all coupled bits in its associated copy-set are immediately appended to the scheduling sequence $\mathcal{S}$ (Lines~\ref{alg:line:copy}--\ref{alg:line:copied}). This mechanism ensures that these bits are treated as known frozen bits in all subsequent reliability evaluations, $Z_N(u_k | \mathcal{S})$, effectively reducing the code rates of the corresponding subblocks.

\begin{algorithm}
\caption{Channel-Aware Scheduling with Parity Checks}
\label{alg:pc_greedy_scheduling}
\begin{algorithmic}[1]
\STATE \textbf{Input:} Information set $\mathcal{I}$, Channel condition $\boldsymbol{\epsilon}$, Copy-sets $\mathcal{C}$
\STATE \textbf{Output:} Parity-aware decoding schedule $\mathcal{S}$

\STATE $\mathcal{S} \gets ()$ \COMMENT{Initialize as an empty ordered sequence}
\STATE $\mathcal{U} \gets \mathcal{I}$ \COMMENT{Initialize set of unscheduled bits}

\WHILE{$\mathcal{U}$ is not empty}
    \STATE \COMMENT{Find the bit with the lowest conditional error probability}
    \STATE $s^* \gets \arg \min_{k \in \mathcal{U}} Z_N(u_k | \mathcal{S})$

    \STATE \COMMENT{Append the selected bit and remove from unscheduled set}
    \STATE Append $s^*$ to $\mathcal{S}$
    \STATE $\mathcal{U} \gets \mathcal{U} \setminus \{s^*\}$

    \STATE \COMMENT{Immediately resolve all associated bit copies} \label{alg:line:copy}
    \FORALL{$w \in \text{CopySet}(s^*, \mathcal{C})$}
        \IF{$w \in \mathcal{U}$}
            \STATE Append $w$ to $\mathcal{S}$
            \STATE $\mathcal{U} \gets \mathcal{U} \setminus \{w\}$
        \ENDIF
    \ENDFOR \label{alg:line:copied}
\ENDWHILE
\RETURN $\mathcal{S}$
\end{algorithmic}
\end{algorithm}

\subsection{Dynamic Interplay: Information Bit Mapping and Decoding Scheduling}\label{section:application:mapping}
A pivotal feature of this framework is the dynamic synergy between the one-to-many information mapping and the arbitrary SC schedule. This interplay is best understood through the following conceptual categorization of subblocks:
\begin{itemize}
    \item \textbf{Capacity-sufficient:} A subblock (typically the mother block) that is fully transmitted with relatively low effective code rate.
    \item \textbf{Capacity-deficient:} A subblock (typically the extension block) that is severely punctured, resulting in an effective code rate that exceeds the channel capacity.
\end{itemize}

Because Algorithm~\ref{alg:generalized_sc_with_pc} instantly freezes copied bits across all subblocks, the effective code rate of each subblock is not static. As the decoder progresses, a capacity-deficient subblock can dynamically become capacity-sufficient.

\begin{figure}
    \centering
    \includegraphics[width=\linewidth]{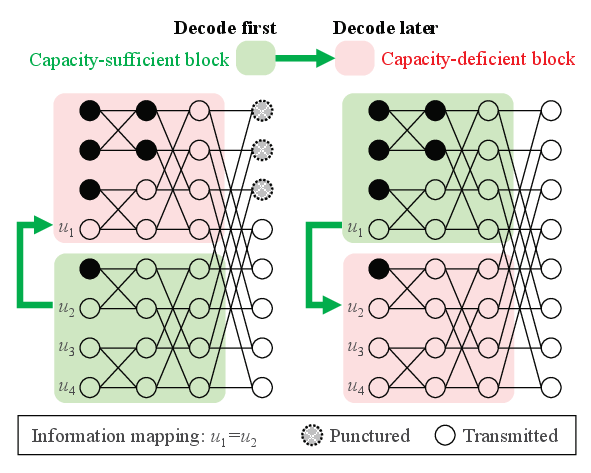}
    \caption{Conceptual categorization of subblocks in terms of rate and capacity.}
    \label{fig:rate-capacity}
\end{figure}

The scheduling algorithm ensures that a capacity-sufficient subblock is decoded first. Once decoded, its decisions assist the capacity-deficient subblocks through freezing the copied bits in those subblocks, as shown in Fig.~\ref{fig:rate-capacity}.

However, to maximize this assistive effect across arbitrary truncations of $E$, the specific mapping order of $u_{\mathcal{I}_q} = u_{\mathcal{I}_p}$ is critical.

Assume the mother subblock (containing $\mathcal{I}_p$) is fully transmitted, while the extension subblock (containing $\mathcal{I}_q$) is partially transmitted due to puncturing. Since the transmitted length $E$ can be any value, we implement a \textit{reverse mapping} strategy: the earliest decoded bits in $\mathcal{I}_p$ are mapped to the earliest transmitted bits in $\mathcal{I}_q$.

For a code of length $2N$ with sequential puncturing (indices $1, \dots, 2N-E$ are punctured), the initial transmission uses indices $N+1, \dots, 2N$. The retransmission uses indices starting from $N$ counting backwards ($N, N-1, \dots, 2N-E+1$). The mapping rule is formalized in Algorithm~\ref{alg:bit_mapping}.

\begin{algorithm}
\caption{Reverse Information Bit Mapping for Rateless IR-HARQ}
\label{alg:bit_mapping}
\begin{algorithmic}[1]
\REQUIRE Sets $\mathcal{I}_p$ and $\mathcal{I}_q$ from nested construction
\ENSURE Bit mapping pairs

\STATE $\mathcal{I}_p^{\text{sort}} \gets \text{Sort } \mathcal{I}_p \text{ by ascending bit index}$
\STATE $\mathcal{I}_q^{\text{sort}} \gets \text{Sort } \mathcal{I}_q \text{ by descending bit index}$ \COMMENT{Aligns with sequential puncturing}
\FOR{$i = 1$ \TO $|\mathcal{I}_p|$}
\STATE Map information bit: $u_{\mathcal{I}_q^{\text{sort}}[i]} \gets u_{\mathcal{I}_p^{\text{sort}}[i]}$
\ENDFOR
\end{algorithmic}
\end{algorithm}

This reverse mapping guarantees that for any transmitted length $E \in [N+1, 2N]$, the capacity-deficient extension block benefits from the earliest decoded bits of the mother block. Specifically, the first decoded bit $u_{s_1}$ in the mother block reduces the effective code rate of the extension block regardless of $E$, because the mapping $u_{s_2} \gets u_{s_1}$ ensures the target index $s_2=N$ is always included in the set of transmitted indices $\{N, N-1, \dots, 2N-E+1\}$.

\begin{remark}
The primary advantage of this dynamic scheduling mechanism is that the algorithm identifies capacity-deficient subblocks as transient bottlenecks. Rather than attempting to decode it prematurely, the decoder waits until the coupled bits in the capacity-sufficient subblock are confidently decided. This ensures an adaptive, on-the-fly matching between rate and capacity.
\end{remark}

\subsection{Simulation results}\label{section:application:simulation}
In this section, we evaluate the performance of the proposed rateless polar codes, implemented using the integrated framework of nested construction, parity-check-adapted scheduling and decoding, and reverse bit mapping (Algorithms~\ref{alg:nested_construction}--\ref{alg:bit_mapping}).

We first examine an exemplary scenario with information length $K=448$ (including a 16-bit CRC) and a mother code range of $N_{\min} = 512$ to $N_{\max} = 1024$. The actual transmitted block length $N$ varies continuously within this interval. Decoding is performed using CRC-aided SCL decoding with a list size $L=8$. The performance metric is the required SNR ($E_s/N_0$) to achieve a target BLER of $10^{-2}$.

\begin{figure}
    \centering
    \includegraphics[width=\linewidth]{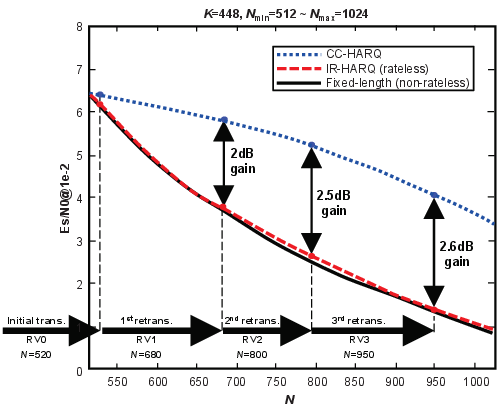}
    \caption{Required $E_s/N_0$ to achieve BLER $= 10^{-2}$ for the proposed rateless polar code ($K=448$) across continuous code lengths $N \in [512, 1024]$.}
    \label{fig:K448N512-1024}
\end{figure}

As illustrated in Fig.~\ref{fig:K448N512-1024}, we compare our proposed scheme against two distinct benchmarks:
\begin{itemize}
    \item \textbf{Chase Combining HARQ (CC-HARQ):} A baseline where redundancy is achieved via simple bit repetition. This scheme provides energy gain but lacks additional coding gain, serving as a lower bound for performance.
    \item \textbf{Fixed-Length QUP Polar Codes:} A benchmark where a unique polar code is constructed and encoded for each specific $N$ using Quasi-Uniform Puncturing (QUP)~\cite{Polar:QUP}. Since these codes are optimized for a single rate and are not restricted by nested constraints, they represent the performance upper bound.
\end{itemize}

To emulate a realistic HARQ process, we examine four specific redundancy versions (RVs) at $N \in \{530, 680, 800, 950\}$. Our observations are as follows:
\begin{enumerate}
    \item \textbf{Significant Coding Gain:} Compared to CC-HARQ, the proposed scheme achieves gains of 0.2~dB, 2.0~dB, 2.5~dB, and 2.6~dB at the respective RVs.
    \item \textbf{Optimality:} Remarkably, the proposed rateless codes achieve similar performance to the fixed-length QUP codes across the entire range. This indicates that the nested constraints do not incur a performance penalty when adaptive decoding schedule is employed.
    \item \textbf{Monotonicity:} The $N$-to-SNR curve is smooth and monotonically decreasing. This behavior is desirable for practical base station scheduling, as it ensures that additional redundancy consistently translates into predictable reliability gains.
\end{enumerate}

\begin{figure*}[!t]
    \centering
    \includegraphics[trim=50mm 20mm 50mm 20mm, clip, width=\linewidth]{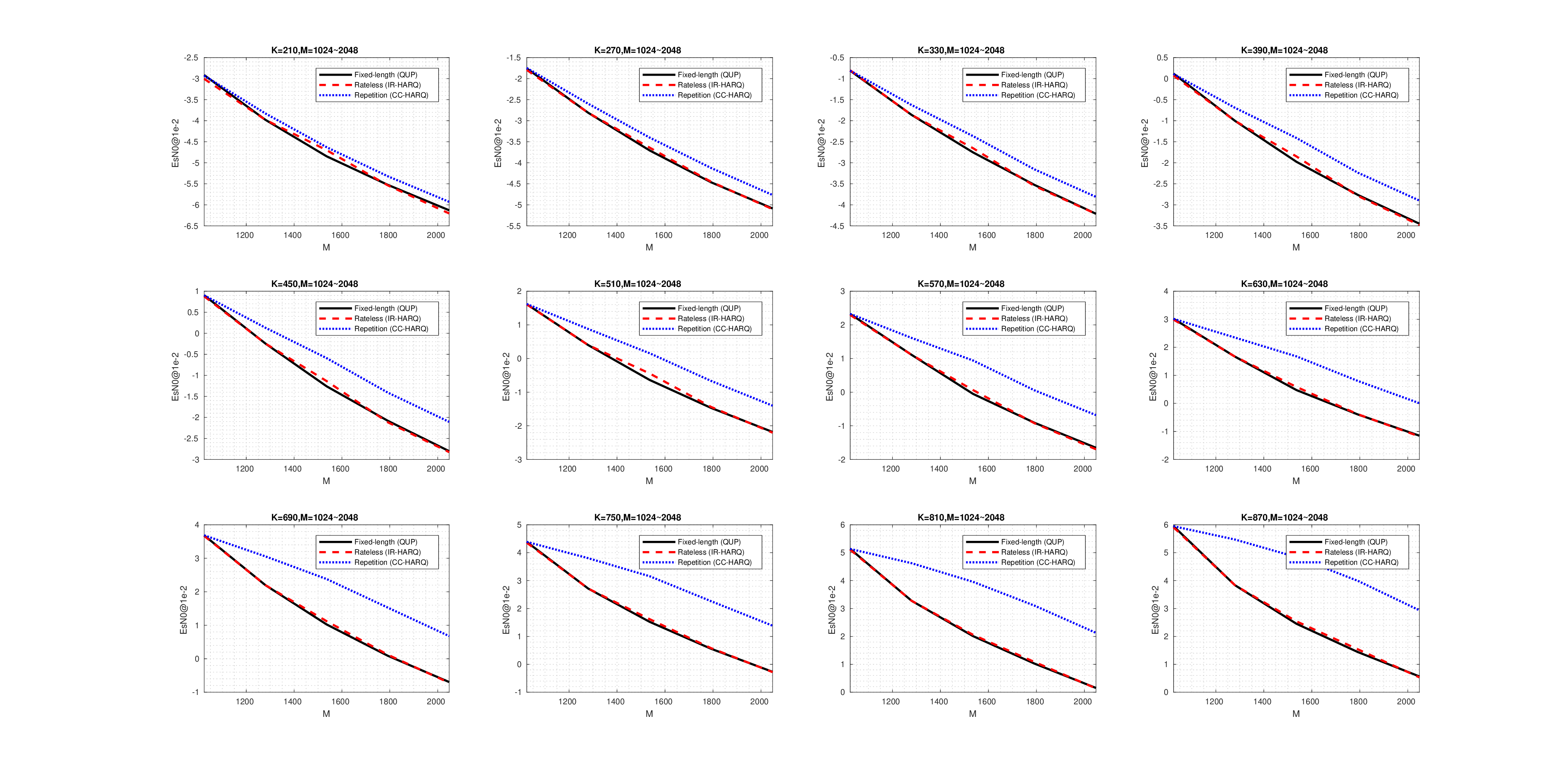}
    \caption{Required $E_s/N_0$ to achieve BLER $= 10^{-2}$ for the proposed rateless polar codes across continuous information lengths $K \in \{210, \dots, 870\}$ and code lengths $N \in [1024, 2048]$.}
    \label{fig:performance}
\end{figure*}

To verify the framework's robustness, we simulated fine-grained performance in a broader range of code rates and lengths, specifically $K \in \{210, \dots, 870\}$ with $N_{\min} = 1024$ and $N_{\max} = 2048$. As shown in Fig.~\ref{fig:performance}, across all simulated rate-length combinations, the proposed scheme consistently matches the performance of independently designed, non-rateless polar codes.

\subsection{Hardware implementation results}
To evaluate the physical feasibility and hardware overhead of the proposed framework, we developed an ASIC (Application-Specific Integrated Circuit) implementation comparing the rateless polar decoder against a conventional polar decoder architecture. To ensure a fair comparison, both decoders are designed with a list size of $L=8$, support a maximum code length of $N_{\max} = 1024$, and were synthesized using the same process node.

\begin{figure}
    \centering
    \includegraphics[width=\linewidth]{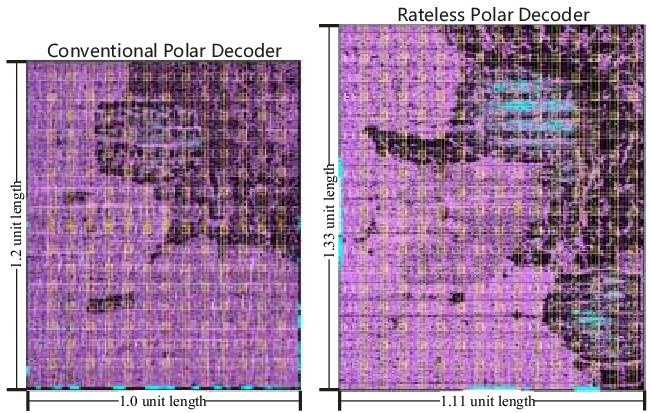}
    \caption{Physical layout plots of the conventional polar decoder and the proposed rateless polar decoder, drawn to the same scale.}
    \label{fig:layout}
\end{figure}

The physical layouts of both decoders are illustrated in Fig.~\ref{fig:layout}. For the purposes of comparison, the chip area of the conventional polar decoder is normalized to $1.00 \times 1.20$ units. Under the same scaling, the proposed rateless polar decoder occupies a footprint of $1.11 \times 1.33$ units, representing an area overhead of approximately 23\%. This modest increase is primarily attributed to the additional control logic and memory required to manage the arbitrary scheduling and parity-check constraints across the decoding list.

The energy efficiency of the two designs was evaluated at a decoding clock frequency of 1~GHz. Power consumption measurements for a $K=256, N=768$ polar code indicate that the rateless polar decoder consumes approximately 1.22$\times$ the power of the conventional baseline. This 22\% power increment is associated with the increased switching activity in the scheduling logic necessitated by providing a unified solution for nested code lengths.

\section{Discussions}
\subsection{Related works}
Adapting polar codes for HARQ systems is fundamentally challenging due to their rigid code-length constraints. While various schemes have been proposed in the literature, they generally fall short of providing a fully flexible, high-performance rateless solution.

Early attempts to implement HARQ for polar codes often relied on CC-HARQ or the transmission of non-polar-coded bits. For instance, the authors in~\cite{Polar:CC-HARQ:EPP} proposed an improved CC-HARQ scheme based on the equivalent puncturing patterns (EPPs) to complement the LLRs of punctured bits. To introduce incremental redundancy, the scheme in \cite{Polar:X-HARQ:info} selectively transmits uncoded information bits as redundancy. This concept was later generalized in~\cite{Polar:X-HARQ:intermediate} by transmitting intermediate bits within the SC trellis. However, because the additionally transmitted bits in these schemes are not fully protected by the extended polar transformation, they fail to enjoy the full coding gain of polar codes.

To achieve optimal coding gain, subsequent research focused on structurally extending the polar code itself. For practical finite block lengths, IR-HARQ schemes based on polarizing matrix extension~\cite{Polar:IR-HARQ,Polar:IR-HARQ2,Polar:IR-HARQ3} generally outperform IF-HARQ by providing extra coding gain. The ``copy bits pair'' mechanism was proposed in~\cite{Polar:IR-HARQ2}, where the value of a bit in the previous transmission is copied to its corresponding paired bit in the extension block. This concept was also combined with quasi-uniform puncturing (QUP) in~\cite{Polar:IR-HARQ3}.

Despite their improved finite-block-length performance, these matrix extension schemes suffer from severe structural limitations. Specifically, the scheme in~\cite{Polar:IR-HARQ2} strictly requires the length of the retransmitted blocks to grow in multiples of two (i.e., the total transmitted length must be a power of two). Similarly, \cite{Polar:IR-HARQ3} is mathematically equivalent to~\cite{Polar:IR-HARQ2} when the incremental lengths follow rigid power-of-two constraints. Even the adaptive extension scheme in~\cite{Polar:IR-HARQ}, which we extensively analyzed in previous sections, lacks a unified decoding scheduling mechanism to gracefully handle arbitrary transmission lengths.

Consequently, none of the existing literature provides a seamless, rateless IR-HARQ framework that maintains near-optimal coding gain for completely arbitrary and continuous code lengths, which motivates the joint code-construction and scheduling design proposed in this paper.

\subsection{Open problems}
The framework presented in this paper introduces a new multi-dimensional design space for polar codes, as illustrated by the additional degrees of freedom in Fig.~\ref{fig:new-DoF}. While we have utilized greedy scheduling (Algorithm~\ref{alg:pc_greedy_scheduling}) and heuristic bit mapping (Algorithm~\ref{alg:bit_mapping}) for their practical efficiency, several fundamental questions remain open regarding the theoretical limits and architectural optimizations of this approach.

A primary open question is the characterization of the \textit{jointly optimal} code construction and decoding schedule. While our greedy approach yields near-optimal coding gains for finite block lengths, a rigorous information-theoretic analysis is required to determine the optimal joint code construction and scheduling strategy that minimizes the required SNR across all possible lengths $E$.

Hardware implementation of the proposed framework for high throughput remains a valuable research direction. Standard polar decoding enhancements, such as simplified SC (SSC)~\cite{Polar:SSC:Kschischang} and Fast-SSCL \cite{Polar:SSCL:Gross}, rely on parallelizing specific node within an SC tree. Extending these techniques to support the \textit{arbitrary} schedules and dynamic parity-check constraints of our framework is non-trivial but essential. Notably, efficient hardware architectures have been developed for existing matrix-extension IR-HARQ schemes in~\cite{Polar:SC-HARQ:Gross,Polar:SCL-HARQ:Gross}. Generalizing these architectures for arbitrary decoding orders and dynamic scheduling offers a pathway toward high-throughput rateless polar coding.

From an application perspective, the channel-aware adaptivity of these rateless polar codes provides a robust mechanism against fading and interference. In such scenarios, specific code bits are subjected to deep fading or erasure. Our framework allows the decoder to strategically circumvent these unreliable bit positions by re-ordering the decoding schedule. A thorough investigation of this capability under standardized fading models (e.g., Rayleigh or Rician) and practical 3GPP preemption patterns could provide valuable insights into designing more resilient communication links for mission-critical applications.

\section{Conclusion}\label{section:conclusion}
In this paper, we have presented a unified framework for rateless polar codes by introducing a new degree of freedom in the design space, i.e., the joint optimization of nested code construction and decoding scheduling. The framework supports arbitrary and nested code lengths $E \in [N_{\min}, N_{\max}]$.
Simulation results demonstrate that the proposed rateless codes achieve near-optimal coding gains, matching the performance of independently optimized, fixed-rate benchmarks across a wide range of rates and lengths. Moreover, an ASIC implementation for $L=8$ and $N_{\max}=1024$ confirms hardware feasibility, offering a robust, channel-aware solution for practical implementations.

\appendices
\section{Proof for the generalized single-step transform}\label{section:appendix:general-transform}
Consider two independent binary-input discrete memoryless channels (B-DMCs) $W_1: \mathcal{X} \to \mathcal{Y}_1$ and $W_2: \mathcal{X} \to \mathcal{Y}_2$. The single-step polarization transform $(W_1, W_2) \to (W', W'')$ is defined via the mapping $X_1 = U_1 \oplus U_2$ and $X_2 = U_2$.

The synthesized transition probabilities are:
\begin{equation}
W'(y_1, y_2 | u_1) = \sum_{u_2 \in \{0,1\}} \frac{1}{2} W_1(y_1 | u_1 \oplus u_2) W_2(y_2 | u_2)
\end{equation}
\begin{equation}
W''(y_1, y_2, u_1 | u_2) = \frac{1}{2} W_1(y_1 | u_1 \oplus u_2) W_2(y_2 | u_2)
\end{equation}

\subsection{Symmetric Capacity Conservation}
By the chain rule of mutual information and the fact that $(U_1, U_2) \to (X_1, X_2)$ is a bijection:
\begin{align*}
I(W_1) + I(W_2) &= I(X_1; Y_1) + I(X_2; Y_2) \\
&= I(X_1, X_2; Y_1, Y_2) \\
&= I(U_1, U_2; Y_1, Y_2) \\
&= I(U_1; Y_1, Y_2) + I(U_2; Y_1, Y_2 | U_1) \\
&= I(W') + I(W'')
\end{align*}
Note: For the specific case where $W_1, W_2$ are Binary Erasure Channels (BECs), $I(W') = I(W_1)I(W_2)$ and $I(W'') = I(W_1) + I(W_2) - I(W_1)I(W_2)$.

\subsection{Reliability of the Enhanced Channel $W''$}
The Bhattacharyya parameter $Z(W'')$ is defined as:
\begin{equation*}
Z(W'') = \sum_{y_1, y_2, u_1} \sqrt{W''(y_1, y_2, u_1 | 0) W''(y_1, y_2, u_1 | 1)}
\end{equation*}

Substituting the definition of $W''$, we obtain $Z(W'')=Z(W_1)Z(W_2)$ (see detailed derivation in \eqref{eq:Z_het_plus_proof}).
\begin{figure*}[!t] 
\begin{align}
Z(W'') &= \sum_{y_1, y_2, u_1} \sqrt{\frac{1}{2} W_1(y_1 | u_1) W_2(y_2 | 0) \cdot \frac{1}{2} W_1(y_1 | u_1 \oplus 1) W_2(y_2 | 1)} \nonumber \\
&= \left( \sum_{u_1 \in \{0,1\}} \frac{1}{2} \right) \left( \sum_{y_1} \sqrt{W_1(y_1 | 0) W_1(y_1 | 1)} \right) \left( \sum_{y_2} \sqrt{W_2(y_2 | 0) W_2(y_2 | 1)} \right) \nonumber \\
&= 1 \cdot Z(W_1) \cdot Z(W_2) = Z(W_1)Z(W_2) \label{eq:Z_het_plus_proof}
\end{align}
\end{figure*}

\subsection{Upper Bound for the Degraded Channel $W'$}
Let $a = W_1(y_1 | 0), b = W_1(y_1 | 1), c = W_2(y_2 | 0), d = W_2(y_2 | 1)$.
\begin{equation*}
Z(W') = \sum_{y_1, y_2} \frac{1}{2} \sqrt{(ac + bd)(ad + bc)}
\end{equation*}
Using the inequality $\sqrt{A^2 + B^2 - C^2} \le A + B - C$ where $A = (a+b)\sqrt{cd}$, $B = (c+d)\sqrt{ab}$, and $C = 2\sqrt{abcd}$, we obtain $Z(W')=Z(W_1) + Z(W_2) - Z(W_1)Z(W_2)$ (see detailed derivation in \eqref{eq:Z_het_minus_proof}).
\begin{figure*}[!t] 
\begin{align}
Z(W') &\le \sum_{y_1, y_2} \frac{1}{2} \left[ (a+b)\sqrt{cd} + (c+d)\sqrt{ab} - 2\sqrt{abcd} \right] \nonumber \\
&= \frac{1}{2} \sum_{y_1} (a+b) \sum_{y_2} \sqrt{cd} + \frac{1}{2} \sum_{y_2} (c+d) \sum_{y_1} \sqrt{ab} - \sum_{y_1} \sqrt{ab} \sum_{y_2} \sqrt{cd} \nonumber \\
&= 1 \cdot Z(W_2) + 1 \cdot Z(W_1) - Z(W_1)Z(W_2) \nonumber \\
&= Z(W_1) + Z(W_2) - Z(W_1)Z(W_2) \label{eq:Z_het_minus_proof}
\end{align}
\end{figure*}
Equality holds if and only if $W_1$ and $W_2$ are BECs.

\section{Proof for channel-dependent decoding scheduling}\label{section:appendix:channel-dependent-scheduling-proof}
In this appendix, we provide the rigorous derivation of subchannel reliabilities and block error probability comparisons for the remaining cases $N=7$ and $N=8$. These derivations substantiate the claim that the optimal decoding schedule is sensitive to the underlying channel parameter $\varepsilon$.

\subsection*{Case 3: Code Length $N=7$}

For $N=7$, we consider a scenario where the first code bit $x_1$ is punctured. The initial channel reliabilities are defined as $Z(x_1)=1$ and $Z(x_i)=\varepsilon$ for $i \in \{2, 3, \dots, 8\}$.

\begin{itemize}
    \item \textbf{Schedule $S_1$ ($u_1 \to u_2 \to u_3 \to u_4$):}
    By applying the generalized recursive relations for heterogeneous inputs, the polarized subchannel reliabilities are obtained as:
    \begin{align*}
        Z_7(u_1) &= (2\varepsilon - \varepsilon^2)^3 \\
        Z_7(u_2|u_1) &= (\varepsilon + \varepsilon^2 - \varepsilon^3)(2\varepsilon^2 - \varepsilon^4) \\
        Z_7(u_3|u_1, u_2) &= \varepsilon^3 + \varepsilon^4 - \varepsilon^7 \\
        Z_7(u_4|u_1, u_2, u_3) &= \varepsilon^7
    \end{align*}
    The resulting block error probability is bounded by $P_B(S_1) \approx 1 - \prod_{i=1}^4 (1 - Z_7(u_i|\dots))$.

    \item \textbf{Schedule $S_2$ ($u_2 \to u_3 \to u_4 \to u_1$):}
    Under the permuted decoding order, the subchannel reliabilities are obtained by
    \begin{align*}
        Z_7(u_2) &= \varepsilon^2(2 - \varepsilon)^2 \\
        Z_7(u_3|u_2) &= 2\varepsilon^2 - \varepsilon^4 \\
        Z_7(u_4|u_2, u_3) &= \varepsilon^4 \\
        Z_7(u_1|u_2, u_3, u_4) &= \varepsilon^3
    \end{align*}
    The resulting block error probability is $P_B(S_2) \approx 1 - \prod_{i \in \{2,3,4,1\}} (1 - Z_7(u_i|\dots))$.

    \item \textbf{Comparison and Threshold Analysis:}
    \begin{proof}
    The difference in block error probabilities $\Delta Z_7 = P_B(S_1) - P_B(S_2)$ can be expressed as:
    \begin{align*}
        &\Delta Z_7=\\
        & -\varepsilon^2(\varepsilon-1)^6(\varepsilon+1)^2(\varepsilon^2+1)(\varepsilon^2+\varepsilon+1) \cdot g(\varepsilon) \\
        &= (\text{Negative}) \times g(\varepsilon)
    \end{align*}
    where $g(\varepsilon)$ is a polynomial of the channel parameter. Evaluating at the boundaries, we find $g(0) = 6 > 0$ and $g(1) = -4 < 0$, and $g(\varepsilon)$ has a unique real root $\varepsilon_{th} \approx 0.920$ in the interval $(0, 1)$.

    Consequently, for $\varepsilon < 0.920$, $\Delta Z_7 < 0$, rendering $S_1$ the superior schedule. For $\varepsilon > 0.920$, the sign flips, and $S_2$ becomes optimal.
    \end{proof}
\end{itemize}

\subsection*{Case 4: Code Length $N=8$}

In the non-punctured case ($N=8$), all input channels are homogeneous with $Z(x_i) = \varepsilon$ for all $i$.

\begin{itemize}
    \item \textbf{Schedule $S_1$ ($u_1 \to u_2 \to u_3 \to u_4$):}
    The standard Ar{\i}kan recursion for the selected indices in $\mathcal{I}$ yields:
    \begin{align*}
        Z_8(u_1) &= (2\varepsilon - \varepsilon^2)^4 \\
        Z_8(u_2|u_1) &= (2\varepsilon^2 - \varepsilon^4)^2 \\
        Z_8(u_3|u_1, u_2) &= 2\varepsilon^4 - \varepsilon^8 \\
        Z_8(u_4|u_1, u_2, u_3) &= \varepsilon^8
    \end{align*}

    \item \textbf{Schedule $S_2$ ($u_2 \to u_3 \to u_4 \to u_1$):}
    Changing the decoding schedule results in:
    \begin{align*}
        Z_8(u_2) &= \varepsilon^2(2 - \varepsilon)^2 \\
        Z_8(u_3|u_2) &= 2\varepsilon^2 - \varepsilon^4 \\
        Z_8(u_4|u_2, u_3) &= \varepsilon^4 \\
        Z_8(u_1|u_2, u_3, u_4) &= \varepsilon^4
    \end{align*}

    \item \textbf{Comparison and Threshold Analysis:}
    \begin{proof}
    The difference $\Delta Z_8 = P_B(S_1) - P_B(S_2)$ is given by:
    \begin{align*}
        &\Delta Z_8 =\\
        & \varepsilon^2(\varepsilon-1)^6(\varepsilon+1)^4(\varepsilon^2+1)^2(\varepsilon^2-2\varepsilon-1) \cdot h(\varepsilon) \\
        &= (\text{Negative}) \times h(\varepsilon)
    \end{align*}
    where $h(\varepsilon)$ is the decision polynomial. We observe $h(0) = 6 > 0$ and $h(1) = -1 < 0$. The polynomial possesses exactly one real root $\varepsilon_{th} \approx 0.965$ within the interval $(0, 1)$.

    For a reliable channel ($\varepsilon < 0.965$), $h(\varepsilon)$ is positive, making $\Delta Z_8 < 0$. Thus, the sequential schedule $S_1$ is optimal. As the channel degrades toward the threshold $\varepsilon_{th} \approx 0.965$, the optimal decoding order switches to $S_2$.
    \end{proof}
\end{itemize}

\section*{Acknowledgment}
The authors would like to express their sincere gratitude to Prof. Erdal Ar{\i}kan for his valuable suggestions and constructive feedback during the preparation of this manuscript. His insights were instrumental in improving the quality and clarity of this work.

\ifCLASSOPTIONcaptionsoff
  \newpage
\fi

\end{document}